\documentclass[12pt,reqno]{amsart}
\usepackage{amssymb}
\usepackage{graphicx}
\usepackage{enumerate}
\usepackage{algorithmic,algorithm}
\usepackage{latexsym, amsmath, amscd, amssymb}
\usepackage{natbib}
\usepackage{setspace}
\usepackage{enumerate}
\usepackage[english]{babel}
\usepackage{fancyhdr}

\newtheorem{lemma}{Lemma}[section]
\newtheorem{theorem}[lemma]{Theorem}
\newtheorem{corollary}[lemma]{Corollary}
\newtheorem{proposition}[lemma]{Proposition}
\newtheorem{definition}[lemma]{Definition}

\newtheorem{example}[lemma]{Example}

\newtheorem{claim*}{Claim}
\begin{document}
\title[Krull Dimension of $A$-algebras]
{On  Gr\"obner Bases and Krull Dimension of Residue Class Rings of Polynomial Rings over Integral Domains}
\author[Maria Francis and  Ambedkar Dukkipati]{Maria Francis and  Ambedkar Dukkipati}

\email{mariaf@csa.iisc.ernet.in \\ ad@csa.iisc.ernet.in}
\address{Dept. of Computer Science \& Automation\\Indian Institute of Science, Bangalore - 560012}
%
\maketitle
%
%
%
\begin{abstract}
Given an ideal $\mathfrak{a}$ in  $A[x_1, \ldots, x_n]$ where $A$ is a Noetherian integral domain,  we propose an approach to compute the Krull dimension of
$A[x_1,\ldots,x_n]/\mathfrak{a}$, when the residue class ring is a free $A$-module. When $A$ is a field,
the Krull dimension of $A[x_1,\ldots,x_n]/\mathfrak{a}$ has several equivalent algorithmic definitions by which it can be computed.
But this is not true in the case of arbitrary Noetherian rings. 
For a Noetherian integral domain $A$ we 
introduce the notion of combinatorial dimension of $A[x_1, \ldots,
  x_n]/\mathfrak{a}$  and give a Gr\"obner basis method to compute it for residue class rings that have a free $A$-module representation w.r.t. a lexicographic ordering. 
For such $A$-algebras, we derive a
relation between Krull dimension and combinatorial dimension of $A[x_1, \ldots,
  x_n]/\mathfrak{a}$.
An immediate application of this
relation is that it gives a uniform method,  the first of its kind, to compute the dimension of
$A[x_1, \ldots, x_n]/\mathfrak{a}$ without having  to consider
individual properties of the ideal.
 For $A$-algebras that have a free $A$-module representation w.r.t. 
degree compatible monomial orderings, we introduce the concepts of Hilbert function, Hilbert series and Hilbert polynomials 
and show that Gr\"obner basis methods can be used to compute these quantities. We then proceed to show that the combinatorial dimension of such $A$-algebras is equal to 
the degree of the Hilbert polynomial.  This enables us to extend the relation between Krull dimension and combinatorial dimension to $A$-algebras with a free $A$-module representation w.r.t. a degree compatible ordering as well.
\end{abstract}

\section{Introduction}
\label{Introduction} 
One of the fundamental problems in computational ideal theory  is determining the dimension of an ideal, i.e. the Krull dimension of the $\Bbbk$-algebra,
$\Bbbk[x_1, \ldots, x_n]/\mathfrak{a}$.
The dimension of an affine variety  
associated with an ideal $\mathfrak{a}\subseteq \Bbbk[x_1, \ldots,
  x_n]$ is the Krull dimension of the affine $\Bbbk$-algebra 
$\Bbbk[x_1, \ldots, x_n]/\mathfrak{a}$ for an algebraically closed field $\Bbbk$.  
 Since the definition of Krull dimension does not lead to an algorithmic
 method to compute it, various alternate equivalent definitions have
 been proposed.  The Krull dimension of an affine $\Bbbk$-algebra is
 equal to its transcendence degree, the degree of the Hilbert polynomial
 of $\mathfrak{a}$ and the largest number   of elements among the maximal
 set  of indeterminates independent mod $\mathfrak{a}$  (called the
 combinatorial dimension of $\Bbbk[x_1, \ldots,   x_n]/\mathfrak{a}$) \citep{Kreuzer:2005:borderbases}.  
Gr\"{o}bner basis based algorithms have been proposed to compute the degree of the
Hilbert polynomial of $\mathfrak{a}$  and the combinatorial dimension
of $\Bbbk[x_1, \ldots, x_n]/\mathfrak{a}$ \citep{mora:1983:hilbertpoly, 
  Kredel:1988:maximalset}, thus providing an algorithmic framework for
determining the Krull dimension of the affine variety associated with
$\mathfrak{a}$. This paper studies the question of whether one can give
Gr\"{o}bner basis methods to compute the Krull dimension of $A[x_1,\ldots,x_n]/\mathfrak{a}$, where $A$ is a Noetherian integral domain, given that it has a free $A$-module representation w.r.t. some monomial order.


For any Noetherian commutative ring $A$,  a necessary and
sufficient condition for a finitely generated $A$-module
$A[x_1,\ldots,x_n]/\mathfrak{a}$ to have a  
free $A$-module representation w.r.t. a monomial order has been studied in \citep{FrancisDukkipati:2013:freeZmodule}. Here, we show that this characterization can be extended to  
$A[x_1,\ldots,x_n]/\mathfrak{a}$ that need not be finitely generated as an $A$-module. 

Given an integral domain $A$ and an $A$-algebra with a free $A$-module representation w.r.t. some monomial order, we study alternate algorithmic definitions for  Krull dimension. 
We first extend the concept of combinatorial dimension to $A$-algebras. 
For an  $A$-algebra with a free $A$-module representation w.r.t a lexicographic ordering, 
we give a Gr\"obner basis algorithm for computing its combinatorial  dimension. 
In affine $\Bbbk$-algebras, the combinatorial dimension is equal to the Krull dimension.
We derive a relation between Krull dimension and combinatorial dimension for $A$-algebras that have a free $A$-module representation w.r.t. 
a lexicographic order.  We also show that the concepts of Hilbert functions, Hilbert series and
 Hilbert polynomial can be extended to  $A$-algebras that have a free $A$-module
 representation w.r.t. a degree compatible ordering. We also give a Gr\"obner basis algorithm to compute these quantities. 
 For  degree
 compatible orderings, we show that the combinatorial dimension of
 $A[x_1,\ldots,x_n]/\mathfrak{a}$   is equal to the degree of the
 Hilbert polynomial of $\mathfrak{a}$ and therefore we have a Gr\"obner basis algorithm to compute the combinatorial dimension. We also show how this can used to derive a  relation between the Krull dimension of $A[x_1,\ldots,x_n]/\mathfrak{a}$  and the degree of the Hilbert polynomial of $\mathfrak{a}$.
The concepts of combinatorial dimension and Hilbert polynomial are important because 
they give us  a uniform method, independent of the ideal, to determine the 
Krull dimension of $A[x_1,\ldots,x_n]/\mathfrak{a}$ that has a free $A$-module representation w.r.t. either a lexicographic or a degree compatible monomial ordering.  More importantly, these concepts allow for an algorithmic interpretation of  the algebraic concept of 
Krull dimension for certain $A$-algebras. 

The rest of the paper is organized as follows.
In Section~\ref{Preliminaries}, we discuss the notations used in the paper.
In Section~\ref{characterization}, we extend the necessary and sufficient condition for the quotient ring $A[x_1,\ldots,x_n]/\mathfrak{a}$, where $A$ is a Noetherian commutative ring,
to have a free $A$-module representation w.r.t. a monomial order 
to the infinite case. 
After this section, the paper restricts its study to residue class rings of polynomial rings over Noetherian integral domains. 
In Section~\ref{dimensionofAalgebra}, we define combinatorial dimension for $A$-algebras, where $A$ is a Noetherian integral domain.  In 
Section~\ref{krulldimensionforlex}, we give a Gr\"obner basis method to compute the combinatorial dimension of $A[x_1,\ldots,x_n]/\mathfrak{a}$ 
that has a free $A$-module representation w.r.t. a lexicographic ordering. 
For such $A$-algebras, we derive a relation between combinatorial dimension and Krull dimension in Section~\ref{dimension}. 
In Section~\ref{exampleslex},  we illustrate with examples how this relation gives an algorithmic method to determine the Krull dimension.
We define Hilbert function, Hilbert series and Hilbert polynomial for $A$-algebras that have a free $A$-module representation w.r.t. a
degree compatible monomial order in Section~\ref{Hilbert}. We also give a Gr\"obner basis algorithm to compute these quantities. 
We then show in Section~\ref{hilbertequalscomb} that the combinatorial dimension of $A$-algebras
with a free $A$-module representation w.r.t. a degree compatible monomial order is equal to the degree of the Hilbert polynomial. 
This enables us to give a relation between 
the degree of the Hilbert polynomial  and the Krull dimension of the corresponding residue class ring in Section~\ref{krullwithhilbertsection}.

\section{Preliminaries}
\label{Preliminaries}
Throughout this paper, $\Bbbk$ denotes a field,   $\mathbb{Z}$ the ring of integers and $\mathbb{N}$ the set of positive integers including zero. 
We use $A$ to denote a Noetherian commutative ring. From Section~\ref{dimensionofAalgebra} onwards, $A$ is restricted to Noetherian integral domains. 
A polynomial ring in indeterminates $x_1,\ldots, x_n$ over $A$ is denoted as $A[x_1,\ldots, x_n]$.  
At times, we represent the indeterminates collectively as a set $X$ and the corresponding polynomial ring  as $A[X]$.
We represent a monomial in $x_1,\ldots, x_n$ as $x^{\alpha}$  where  $\alpha\in {\mathbb{Z}}_{\ge 0}^n$. 
The monoid isomorphism between the set of all monomials in indeterminates $x_1,\ldots, x_n$ and  ${\mathbb{Z}}_{\ge 0}^n$ allows us to denote the set of all monomials as ${\mathbb{Z}}_{\ge 0}^n$.
A nonzero polynomial, $f$ in  $x_{1}, \ldots, x_{n}$ with coefficients from $A$ is given by
        \begin{displaymath}
	 f = \sum_{\alpha \in \Lambda_{f}} a_{\alpha} x^{\alpha} \enspace,
	\end{displaymath}
         where $\Lambda_{f} \subsetneq {\mathbb{Z}}_{\geq 0}^{n}$ is a finite set and $a_{\alpha} \in A\setminus \{0\}$.
We denote all the monomials of a polynomial $f$ as $\mathrm{Mon}(f)$.
We assume that there is a monomial order $\prec$ on the monomials in the indeterminates $x_{1},\ldots,x_{n}$. 
With respect to this monomial order, we have the leading monomial ($\mathrm{lm}_\prec$), leading coefficient ($\mathrm{lc}_\prec$), leading term ($\mathrm{lt}_\prec$) and 
multidegree ($\mathrm{multideg}_\prec$) of a polynomial $f\in A[x_1, \ldots, x_n]$, 
where $\mathrm{multideg}_\prec(f) = \mathrm{max}_{\prec}\{\alpha \in \Lambda_{f} \}$  and $\mathrm{lt}_\prec(f) = \mathrm{lc}_\prec(f)\mathrm{lm}_\prec(f)$ in $A[x_{1},\ldots,x_{n}]$. 
In certain scenarios,  we also consider another concept of degree of a polynomial which we will denote  as $\mathrm{deg}(f)$. The degree of a monomial $x^\alpha$, $\mathrm{deg}(x^\alpha)$ is 
the sum of its exponents. 
The degree of a polynomial, $f$  is the maximum degree of the monomials in $f$, i.e.
\begin{displaymath}
\mathrm{deg}(f) = \mathrm{max}\{\mathrm{deg}(x^\alpha): x^\alpha \in \mathrm{Mon}(f)\}.
\end{displaymath}
A degree compatible monomial ordering $\prec$ is a monomial ordering on $A[x_1, \ldots, x_n]$ such that two monomials $x^\alpha, x^{\alpha^{'}}$ with $x^\alpha \prec x^{\alpha^{'}}$ 
satisfy $\mathrm{deg}(x^\alpha) \leq \mathrm{deg}(x^{\alpha^{'}})$. 
For a degree compatible monomial ordering, the leading monomial will be a monomial with maximum degree.
The leading term ideal (or initial ideal) of a set $S \subseteq A[x_{1},\ldots,x_{n}]$, is  $\langle\mathrm{lt}_\prec(S) \rangle = \langle \{\mathrm{lt}_\prec(f) \mid f \in S \} \rangle$. 
When there is no confusion regarding which monomial order to consider we omit the monomial order subscript $\prec$ from the notations.
For a free $A$-module $M$, the minimum cardinality of a basis of $A$ is called its free rank and is denoted by $\mathrm{FreeRank}_A(M)$.
\section{ Characterization of a Free Residue Class Ring of $A[x_1, \ldots, x_n]$}
\label{characterization}
Consider an ideal $\mathfrak{a}$ in $A[x_1,\ldots,x_n]$ and  let $G
= \{g_i: i = 1, \ldots, t\}$  be its Gr\"obner basis w.r.t. a monomial order $\prec$. 
For each monomial $x^\alpha$, let $J_{x^{\alpha}} = \{i : \mathrm{lm}(g_i)\mid x^{\alpha},  g_i \in G \}$  
and $I_{J_{x^{\alpha}}} = \langle \{\mathrm{lc}(g_i) : i \in
J_{x^{\alpha}}\} \rangle$. 
We refer to $I_{J_{x^{\alpha}}}$ as the leading coefficient ideal
w.r.t. $G$.
Let $C_{J_{x^{\alpha}}}$ represent a set of coset representatives of the equivalence classes in  $A/I_{J_{x^{\alpha}}}$.
We  use the same definitions as given in \citep{Adams:1994:introtogrobnerbasis}.
Given a polynomial $f \in  A[x_1,\ldots,x_n]$, let $f =
\sum\limits_{i=1}^m a_i x^{\alpha_i}\hspace{2pt} \mathrm{mod} \hspace{2pt}\langle G \rangle$,
where $a_i \in A, i=1,\ldots,m$.   
If $A[x_{1},\ldots,x_{n}]/\langle G \rangle$ is an  $A$-module finitely generated by $m$ elements,
then corresponding to the coset representatives, $C_{J_{x^{\alpha_1}}}, \ldots, C_{J_{x^{\alpha_m}}}$, 
there exists an $A$-module isomorphism, 
\begin{equation} \label{equation}
\begin{split}
 \phi :  A[x_{1},\ldots,x_{n}]/\langle G \rangle &\longrightarrow A/I_{J_{x^{\alpha_1}}} \times \cdots \times A/I_{J_{x^{\alpha_m}}}\\
         \sum\limits_{i=1}^m a_i x^{\alpha_i} + \langle G \rangle &\longmapsto (c_1 +I_{J_{x^{\alpha_1}}}  , \cdots, c_m + I_{J_{x^{\alpha_m}}}) ,
\end{split}
\end{equation}
where $c_i = a_i \text{  mod  } I_{J_{x^{\alpha_i}}}$ and  $c_i \in C_{J_{x^{\alpha_i}}}$. 

We refer to $A/I_{J_{x^{\alpha_1}}} \times \cdots \times A/I_{J_{x^{\alpha_m}}}$ as the $A$-module representation of 
$A[x_{1},\ldots,x_{n}]/\mathfrak{a}$ w.r.t. $G$ (or w.r.t. $\prec$).  
If $I_{J_{x^{\alpha_i}}} = \{0\} $, we have $C_{J_{x^{\alpha_i}}} = A$,
$\text{for all } i = 1,\ldots, m$.  
This implies  $A[x_{1},\ldots,x_{n}]/\mathfrak{a} \cong A^m$, i.e. $A[x_{1},\ldots,x_{n}]/\mathfrak{a}$ has an $A$-module basis and it is free. 
We say that $A[x_{1},\ldots,x_{n}]/\mathfrak{a} $ has a free $A$-module representation w.r.t. $G$ (or w.r.t. $\prec$). 
If the $A$-module is infinitely generated, we say that it has a free $A$-module representation w.r.t. $G$ (or equivalently w.r.t. $\prec$) if $I_{J_{x^{\alpha}}} = \{0\} $
for all $x^{\alpha} \notin \langle \mathrm{lm}(\mathfrak{a})\rangle$ and $I_{J_{x^{\alpha}}} = A$ for all $x^{\alpha} \in \langle \mathrm{lm}(\mathfrak{a})\rangle$.  

The necessary and sufficient condition for an $A$-module
$A[x_1,\ldots,x_n]/ \mathfrak{a} $ to have a free $A$-module representation w.r.t. $G$ (or w.r.t. $\prec$) makes use of the  the concept of 
`short reduced Gr\"obner basis' introduced in  \citep{FrancisDukkipati:2013:freeZmodule} which we briefly describe below.
\begin{definition}
 Let $\mathfrak{a} \subseteq A[x_1,\ldots,x_n]$ be an ideal.  
 Consider the isomorphism in \eqref{equation}.
 A reduced Gr\"obner basis (as defined in \citep{Pauer:2007:Grobnerbasisrings}), $G$ of $\mathfrak{a}$  is called a short reduced Gr\"obner basis if
 for each $x^\alpha \in \mathrm{lm}(G)$, the number of elements in the generating set of the  leading coefficient ideal of $x^\alpha$, $I_{J_{x^\alpha}}$ in \eqref{equation} is minimal. 
\end{definition}
One can define reduced Gr\"obner bases over rings exactly as that of fields but it may not exist in all the cases. 
The definition of reduced Gr\"obner basis  given by \citep{Pauer:2007:Grobnerbasisrings} ensures the existence of such a basis for every ideal in the polynomial ring. 

A necessary and sufficient condition for a finitely generated $A$-module $A[x_1,\ldots,x_n]/\mathfrak{a}$ to have  a free $A$-module representation 
w.r.t. $G$ (or w.r.t. $\prec$) is given in \citep{FrancisDukkipati:2013:freeZmodule}. 
One can easily extend this to residue class rings that are not finitely generated as shown below. 
\begin{lemma} \label{lemma}
Let $\mathfrak{a} \subseteq A[x_1,\ldots,x_n]$ be a non-zero ideal and let $G$ be a short reduced Gr\"obner basis for $\mathfrak{a}$.
All the leading coefficient ideals associated with $G$ are either trivial or the entire ring $A$ if and only if $G$ is monic. 
\end{lemma}
\proof
The proof is along the lines of \citep[Lemma 3.10]{FrancisDukkipati:2013:freeZmodule}.
\endproof
We now prove the necessary condition.
\begin{theorem}
Let $\mathfrak{a} \subseteq A[x_1,\ldots,x_n]$ be a non-zero ideal and let $G$ be a short reduced Gr\"obner basis of $\mathfrak{a}$. 
If  $A[x_1,\ldots,x_n]/\mathfrak{a}$ has a free $A$-module representation w.r.t. $G$ (or w.r.t. $\prec$) then $G$ is monic. 
\end{theorem}
\proof
By definition, if  $A[x_1,\ldots,x_n]/\mathfrak{a}$ has a free $A$-module representation w.r.t. $G$ then  $I_{J_{x^{\alpha}}} = \{0\} $
for all $x^{\alpha} \notin \langle \mathrm{lm}(\mathfrak{a})\rangle$ and $I_{J_{x^{\alpha}}} = \{1\} $ for all $x^{\alpha} \in \langle \mathrm{lm}(\mathfrak{a})\rangle$.  
That is,  all the leading coefficient ideals associated with $G$ are either trivial or the entire ring, $A$.  Therefore by Lemma~\ref{lemma}, $G$ is monic.
\endproof
The sufficient condition is subsumed by \citep[Theorem 3.8]{FrancisDukkipati:2013:freeZmodule}. 
We state the characterization result as follows. 
\begin{proposition}\label{Proposition for characterization}
Let $\mathfrak{a} \subseteq A[x_1,\ldots,x_n]$ be a nonzero ideal.
Let $G$ be a  short reduced Gr\"obner basis for $\mathfrak{a}$ w.r.t. some monomial ordering $\prec$. Then, $A[x_1,\ldots,x_n]/\mathfrak{a}$ has a free $A$-module representation w.r.t. $G$ (or $\prec$)
if and only if $G$ is monic.
\end{proposition}
\begin{example}
 Let $G = \{3x,5x\}$ be the Gr\"obner basis of an ideal  $\mathfrak{a} $ in $\mathbb{Z}[x,y]$ w.r.t. a lexicographic ordering $\prec$ such that $x \prec y$. 
 We have, $I_{J_{x}} = \langle 3,5 \rangle = \langle 1 \rangle$. Therefore, $\mathbb{Z}[x,y]/\langle 3x,5x \rangle$ has a free $A$-module representation w.r.t. $\prec$.
 The short  reduced Gr\"obner basis of  $\mathfrak{a}$ is given by $G_\mathrm{red} = \{x\}$.
It is monic and a $\mathbb{Z}$-module basis of  $\mathbb{Z}[x,y]/\mathfrak{a}$ is given by $\{1+\mathfrak{a}, y^n+\mathfrak{a}, n \in \mathbb{N} \setminus \{0\}\}$. 
\end{example}
Note that if a Gr\"obner basis or reduced Gr\"obner basis of an ideal w.r.t. a monomial order $\prec$ is monic its short reduced Gr\"obner basis w.r.t. $\prec$ will also be monic, but not vice-versa. Throughout this paper, we will assume that $A[x_1,\ldots,x_n]/\mathfrak{a}$ has a free $A$-module representation w.r.t. a monomial order or equivalently, 
w.r.t. any Gr\"obner basis basis corresponding to that monomial order. 

 \section{Combinatorial dimension of $A[x_1, \ldots, x_n]/\mathfrak{a}$ } 
 \label{dimensionofAalgebra}
 In the rest of the paper we assume that $A$ is a Noetherian integral domain.
The Krull dimension of a ring is defined as the  supremum of the lengths of all the chains of prime ideals in it. 
There are many alternate algorithmic definitions for the dimension of an affine $\Bbbk$-algebra.  
All of them can be shown to be equivalent. On the other hand, for $A$-algebras these definitions are either not equivalent or not valid. 

We define combinatorial dimension of $A[x_1, \ldots, x_n]/\mathfrak{a}$, denoted by $\mathrm{cdim}(A[x_1, \ldots,x_n]/\mathfrak{a})$, in a manner analogous to the definition of 
combinatorial dimension of $\Bbbk[x_1, \ldots, x_n]/\mathfrak{a}$ \citep{Kreuzer:2005:borderbases}.
\begin{definition}
 Given a Noetherian integral domain $A$, let $\mathfrak{a} \subseteq A[x_1, \ldots, x_n]$ be an ideal. Let $X \subseteq \{x_1 ,\ldots, x_n \}$ be a set of indeterminates.
 The set $X$ is said to be independent modulo $\mathfrak{a}$ or an independent
set of indeterminates modulo $\mathfrak{a}$ if $\mathfrak{a} \cap A[X ] = \{0\}$.
 The set $X$ is called a maximal independent set modulo $\mathfrak{a}$ if $X$ is
independent modulo $\mathfrak{a}$ and there is no set $Y \subseteq \{x_1, \ldots, x_n \}$ independent
modulo $\mathfrak{a}$ with $X \subsetneq Y$.
  The largest number of elements of a maximal independent set of indeterminates 
modulo $\mathfrak{a}$ is called the combinatorial dimension of $A[x_1, \ldots, x_n]/\mathfrak{a}$, denoted as $\mathrm{cdim}(A[x_1, \ldots, x_n]/\mathfrak{a})$.                                                                       
\end{definition}
\subsection{Some properties of combinatorial dimension}
The Krull dimension of $A[x_1, \ldots, x_n]/\mathfrak{a}$, for an ideal $\mathfrak{a} \subseteq A[x_1, \ldots, x_n]$, is the maximal Krull dimension of an isolated prime ideal
  associated with $\mathfrak{a}$. 
  Below we show that this result holds for combinatorial dimension as well.
   \begin{lemma}\label{isolatedprimeideal}
 Let $A$ be a Noetherian integral domain and $\mathfrak{a}$ be an ideal in $A[x_1, \ldots, x_n]$. Then $\mathrm{cdim}(A[x_1, \ldots, x_n]/\mathfrak{a})$ is the maximum of 
 $\mathrm{cdim}(A[x_1, \ldots, x_n]/\mathfrak{p})$, where $\mathfrak{p}$ is an isolated prime ideal associated with $\mathfrak{a}$.
  \end{lemma}
\proof
We  will denote  $\mathrm{cdim}(A[x_1, \ldots, x_n]/\mathfrak{a})$ as $d$. 
Let $\mathfrak{p}$ be an isolated prime ideal associated with $\mathfrak{a}$ and $S \subseteq X$ denote the maximal set of indeterminates that are 
independent modulo $\mathfrak{p}$. They are independent modulo $\mathfrak{a}$ and therefore, $d \geq |S|$.  Conversely, let $S \subseteq X$ be a maximal independent
set of indeterminates modulo $\mathfrak{a}$ such that $|S| = d$.  
Then $M = A[S]\setminus \{0\}$ is multiplicatively closed and disjoint from $\mathfrak{a}$. 
There exists a 
prime ideal $\mathfrak{P}$, that contains $\mathfrak{a}$ and does not meet $M$. Let $\mathfrak{p}^{'} \subseteq \mathfrak{P}$ be the isolated prime ideal associated with $\mathfrak{a}$. 
$S$ is 
independent modulo $\mathfrak{p}^{'}$. 
This implies,
 $\mathrm{cdim}(A[x_1, \ldots, x_n]/\mathfrak{p}^{'}) \geq d$.
\endproof

For a subset of indeterminates $S$, the set $\overline{S}$ represents the set of residue classes of $S$ modulo the ideal $\mathfrak{a}$.
 \begin{proposition}\label{samecardinality}
Given a Noetherian integral domain $A$, let $\mathfrak{a} \subseteq A[x_1, \ldots, x_n]$ be a prime ideal. Then, all maximal sets of indeterminates independent modulo $\mathfrak{a}$ have the same 
cardinality. 
\end{proposition}
\proof
Since $\mathfrak{a}$ is a prime ideal, $A[x_1, \ldots, x_n]/\mathfrak{a}$ is an integral domain. 
Let $\mathrm{Quot}(A[x_1, \ldots, x_n]/\mathfrak{a})$ represent the quotient field of $A[x_1, \ldots, x_n]/\mathfrak{a}$.  Let $X = \{x_1, \ldots, x_n\}$ and 
$S\subseteq X$ be a set of indeterminates. $S$ is independent modulo $\mathfrak{a}$ if and only if $\overline{S}$ is algebraically independent
in $\mathrm{Quot}( A[x_1, \ldots, x_n]/\mathfrak{a})$ over $A$.  Assume that there are  maximal independent sets modulo $\mathfrak{a}$  of different cardinalities. 
Let the two sets that are maximal independent modulo $\mathfrak{a}$ be  $S \cup \{a\}$ and $S\cup \{b_1, b_2\}$. This implies $S \cup \{a,b_1\}$ and  $S\cup\{a,b_2\}$ are dependent sets 
of  indeterminates modulo $\mathfrak{a}$. Therefore, we have $\overline{b_1}$ is algebraic over $\mathrm{Quot}(A[\overline{S}])(\overline{a})$ and $\overline{a}$ is
algebraic over $\mathrm{Quot}(A[\overline{S}])(\overline{b_2})$. Therefore, $\overline{b_1}$ is algebraic over $\mathrm{Quot}(A[\overline{S}])(\overline{b_2})$,
which is a contradiction to the independence of $S \cup \{b_1, b_2\}$ modulo $\mathfrak{a}$.
\endproof
\subsection{ Gr\"obner basis method for computing combinatorial dimension for lexicographic orderings }
\label{krulldimensionforlex}
We extend the concept of strongly independent indeterminates modulo $\mathfrak{a}$ introduced in \citep{Kredel:1988:maximalset} for ideals in $\Bbbk[x_1, \ldots, x_n]$, to polynomial rings over $A$. 
\begin{definition}
 Let $S \subseteq X$ be a set of indeterminates and $\prec$ a monomial order  in $A[x_1, \ldots, x_n]$. 
 Then, $A[S/(X\setminus S)]$ denotes the following set,
 \begin{displaymath}
  A[S/(X\setminus S)] = \{f\in A[x_1, \ldots, x_n] : 0 \neq f \text{ and } \mathrm{lt}(f) \in A[S] \}.
 \end{displaymath}
We say that $S$ is strongly independent modulo $\mathfrak{a}$ if $A[S/(X\setminus S)] \cap \mathfrak{a} = \emptyset$. 
\end{definition}
Clearly, if $S$ is strongly independent modulo $\mathfrak{a}$, then it is independent modulo $\mathfrak{a}$. 
But the converse is not true.
\begin{lemma}
Let $\mathfrak{a} \subseteq A[x_1, \ldots, x_n]$ be a proper ideal and $\prec$ be a monomial order 
in $A[x_1, \ldots, x_n]$. 
 Let $S \subseteq X$ be a set of indeterminates.
\begin{enumerate}[(i)]
\item If $S$ is strongly independent modulo $\mathfrak{a}$  w.r.t. $\prec$, then there exists an isolated prime ideal $\mathfrak{p}$ associated with $\mathfrak{a}$ such that 
$S$ is also strongly independent modulo $\mathfrak{p}$ w.r.t. $\prec$.
\item Let $U = \{S \subseteq X : S$  is strongly independent modulo $\mathfrak{a} \text { w.r.t. } \prec \}$ and 
 $U ^{'} = \{S \subseteq X :$ there exists an isolated prime ideal $\mathfrak{p} \text{ associated with } \mathfrak{a} $
  such that $S$ is strongly independent modulo $\mathfrak{p} \text { w.r.t. } \prec \}$, then $U = U^{'}$. 
\end{enumerate}
\end{lemma}
\proof
\begin{enumerate}[(i)]
\item Let $S$ be strongly independent modulo $\mathfrak{a}$. 
Let $M = A[S/(X \setminus S)]\setminus\{0\}$ be a multiplicatively closed subset of $A[x_1, \ldots, x_n]$ disjoint to $\mathfrak{a}$. Then there exists a 
prime ideal $\mathfrak{P}$ such that $\mathfrak{a} \subseteq \mathfrak{P}$ and disjoint from $M$. 
Let $\mathfrak{p}^{'} \subseteq \mathfrak{P}$ be an isolated prime ideal associated with $\mathfrak{a}$.
Then $S$ is strongly independent 
modulo $\mathfrak{p}^{'}$. 
Also, if $S$ is maximal strongly independent modulo $\mathfrak{a}$, then for any $S \subseteq S^{'} \subseteq X$, 
where $S^{'}$ is strongly independent modulo $\mathfrak{p}^{'}$, $S^{'}$ is strongly independent modulo $\mathfrak{a}$, so $S^{'} = S$.
\item Clearly, $U^{'} \subseteq U$ and by (i), $U \subseteq U^{'}$.    
\endproof
\end{enumerate}

We recall the concept of inessential set of indeterminates from \citep{Kredel:1988:maximalset}. Let $S \subseteq X$ be a set of indeterminates, $f \in A[x_1, \ldots, x_n]$ be a polynomial and
$\prec$ be a monomial order in $A[x_1, \ldots, x_n]$. 
We denote $f^{S}$ as the polynomial resulting from $f$ by substituting $1$ for all indeterminates from $S$ in $f$. We say that $S$ is inessential for $f$ if for all 
terms $t$ occurring in $f$, $t^{S} \preccurlyeq \mathrm{lt}(f)^{S}$.
\begin{theorem}\label{inessential}
Let $S \subseteq X$, $\mathfrak{a}$ be a prime ideal in $A[x_1, \ldots, x_n]$ and let $\prec$ a monomial order such that  $A[x_1, \ldots, x_n]/\mathfrak{a}$ has  a free $A$-module
representation w.r.t. $\prec$. Assume that $S$ is independent modulo $\mathfrak{a}$ and that for any $x \in X\setminus S$, there exists 
a polynomial $f_x \in A[S \cup \{x\}/X\setminus (S \cup \{x\})] \cap \mathfrak{a}$ such that $S$ is inessential for $f_{x}$. 
Then $S$ is maximal independent modulo $\mathfrak{a}$ and
$|S| = \mathrm{cdim}(A[x_1,\ldots,x_n]/\mathfrak{a})$. 
\end{theorem}
\proof
For $x \in X\setminus S$, let $d_x$ be the degree of $\mathrm{lt}(f_x)$ in $x$. Then $d_x \gneq 0$, for otherwise $\mathrm{lt}({f_x})^S = 1$ and 
so $t^S = 1$ for all terms $t$ occurring in $f_x$, which implies $f_x \in A[S]$ which contradicts the independence of $S$ modulo $\mathfrak{a}$. 
Let $T$ be the set of all $t \in \mathrm{Mon}(A[x_1, \ldots, x_n])$ such that for every $x \in X \setminus S$, the degree of $x$ in $t$ is $\lneq d_x$.
\begin{claim*}\label{claim1}
For every $t \in \mathrm{Mon}(A[x_1, \ldots, x_n])\setminus T$, there exists $0\neq p, p_1, \ldots, p_m \in A[S]$,
$t_1, \ldots, t_m \in T$ and $f\in \mathfrak{a}$ such that $pt = p_1t_1 + \cdots +p_mt_m+f$.
\end{claim*}
Proof of the claim: Assume the contradiction, that the claim fails for some $t \in \mathrm{Mon}(A[x_1, \ldots, x_n])\setminus T$ and that $t$ is $\prec$-minimal among 
the monomials with this property. Choose $x \in X\setminus S$ such that the degree $d$ of $t$ in $x$ $\geq d_x$ and let $u = tx^{-d} \in \mathrm{Mon}(A[x_1, \ldots, x_n])$. $f_x$ can  
be written as $px^{d_x} - (p_1t_1 + \cdots + p_mt_m)$ with $0\neq p, p_1, \ldots, p_m \in A[S]$, $t_i \in \mathrm{Mon}(A[X\setminus S])$, $t_i \prec x^{d_x}$,
$1\leq i \leq m$. By multiplying with $x^{d-d_x}u$, we get
\begin{displaymath}
  pt = x^{d-d_x}uf_x - (p_1t_1x^{d-d_x}u + \cdots + p_mt_mx^{d-d_x}u).
\end{displaymath}
We have $x^{d-d_x}uf_x \in \mathfrak{a}$ and $t_ix^{d-d_x}u \prec x^{d_x}x^{d-d_x}u = t$ for $1\leq i \leq m$. 
(Note that here we have two comparisons, one is the less than comparison, $<$, based on the degrees of a
 variable in the monomials and the other is the comparison based on the monomial order, $\prec$.) 
Since $t$ is $\prec$-minimal among the monomials that violate the claim, 
the claim is valid for all $t_ix^{d-d_x}u, 1\leq i \leq m$ and therefore Claim~\ref{claim1} is valid for $t$ as well, a contradiction.

Let $\mathrm{Quot}(A)$ represent the quotient field of the integral domain, $A$. Since  $A[x_1, \ldots, x_n]/\mathfrak{a}$ has  a free $A$-module
representation w.r.t. $\prec$ we have $A\cap \mathfrak{a} = \{0\}$. This implies, 
$\mathrm{Quot}(A) \subseteq \mathrm{Quot}(A)(\overline{S}) \subsetneq \mathrm{Quot}(A[x_1, \ldots, x_n]/\mathfrak{a})$. By Claim~\ref{claim1}, 
$\mathrm{Quot}(A)[x_1, \ldots, x_n]/\mathfrak{a}$ is finitely generated as a $\mathrm{Quot}(A)(\overline{S})$-vector space by $\overline{T}$. 
Each $\overline{x}$, $x \in X\setminus S$ is algebraic over $\mathrm{Quot}(A)(\overline{S})$.
Since $A\cap \mathfrak{a} = \{0\}$, this implies that for each $x \in X\setminus S$ we can determine a $f \in A[S \cup \{x\}] \cap \mathfrak{a}$. Therefore, $S$ is maximal independent modulo $\mathfrak{a}$ and 
since $\mathfrak{a}$ is a prime ideal, $\mathrm{cdim}(A[x_1,\ldots,x_n]/\mathfrak{a}) = |S|$. 
\endproof
\begin{definition}[Left Basic Set (LBS)]
Let $\prec$ be a monomial order in $A[x_1, \ldots, x_n]$ and $\mathfrak{a}$ be an ideal in $A[x_1, \ldots, x_n]$. Given the set of indeterminates $X$, 
we define $S_k\subseteq X, 0\leq k \leq n$ inductively as,
\begin{align*}
S_0 &= \emptyset \\ 
 S_{k+1} &=  \begin{cases} 
       S_k \cup \{x_k\}& \text{if } S_k \cup \{x_k\} \text{ is strongly independent } \\& \text{modulo } \mathfrak{a} \text{ w.r.t. } \prec\\
      S_k  & \text{ otherwise. }
   \end{cases}
\end{align*}
The set $S_n$ is called the left basic set of $\mathfrak{a}$ w.r.t. $\prec$. 
\end{definition}
$S_n$ is maximal strongly independent modulo $\mathfrak{a}$ w.r.t. $\prec$. 
For lexicographic orderings, as a consequence of Theorem~\ref{inessential}  we have the following result for prime ideals.
\begin{corollary}
Let $\mathfrak{a}$ be a prime ideal in $A[x_1, \ldots, x_n]$ and $\prec$ be a lexicographic ordering such that 
$A[x_1, \ldots, x_n]/\mathfrak{a}$ has a free $A$-module representation w.r.t. $\prec$. 
If $S$ is the left basic set of $\mathfrak{a}$ w.r.t. $\prec$, then $S$ is maximal independent modulo $\mathfrak{a}$ and so $|S| = \mathrm{cdim}(A[x_1,\ldots,x_n]/\mathfrak{a})$.  
\end{corollary}
\proof
Since $S$ is maximal strongly independent modulo $\mathfrak{a}$, for every $x \in X\setminus S$, there exists a polynomial $f_x \in A[S \cup \{x\}/X\setminus (S\cup \{x\})] \cap \mathfrak{a}$. $f_x$ 
contains no $y \in X$ such that $x \prec y$ since $\prec$ is a lexicographic order. Also for every monomial $t \in \mathrm{Mon}(f_x)$, the degree of $t$ in $x$ is less than or equal to the degree of 
the leading term of $f_x$ in $x$. Therefore, $\mathrm{lt}(f_x)^{S} \geq t^S$ for all terms in $f_x$. Therefore, $S$ is inessential for $f_x$ and we can apply Theorem~\ref{inessential} 
and $|S| = \mathrm{cdim}(A[x_1,\ldots,x_n]/\mathfrak{a})$.   
\endproof
\noindent The idea can be extended to other proper ideals in $A[x_1, \ldots, x_n]$. 
\begin{theorem}\label{dimensionforideals}
Let $\mathfrak{a}$ be a proper ideal in $A[x_1, \ldots,x_n]$ and $\prec$ be a lexicographic monomial order such that $A[x_1, \ldots, x_n]/\mathfrak{a}$ has a free $A$-module representation w.r.t. $\prec$. Let 
\begin{align*}
d = \mathrm{max}&\{|S| : S\subseteq X, S \text{ is maximal strongly } \text{ independent modulo  } \mathfrak{a} \text{ w.r.t. } \prec \}.
\end{align*}
Then, $d = \mathrm{cdim}(A[x_1, \ldots,x_n]/\mathfrak{a})$. 
\end{theorem}
\proof
Since each $S$ that is maximal strongly independent modulo $\mathfrak{a}$ is independent modulo $\mathfrak{a}$ 
we have  $\mathrm{cdim}(A[x_1, \ldots,x_n]/\mathfrak{a})$  $\geq d$. Pick an isolated prime ideal $\mathfrak{p}$  associated with $\mathfrak{a}$ such that 
\begin{displaymath}
\mathrm{cdim}(A[x_1, \ldots, x_n]/\mathfrak{p}) = \mathrm{cdim}(A[x_1, \ldots, x_n]/\mathfrak{a}). 
\end{displaymath}
Let $S$ be the LBS of $\mathfrak{p}$. 
Then, 
\begin{displaymath}
|S| = \mathrm{cdim}(A[x_1, \ldots, x_n]/\mathfrak{p}) = \mathrm{cdim}(A[x_1, \ldots, x_n]/\mathfrak{a}) 
\end{displaymath}
and $S$ is strongly independent modulo $\mathfrak{p}$ and 
therefore  $\mathfrak{a}$ and so $d \geq \mathrm{cdim}(A[x_1, \ldots, x_n]/\mathfrak{a})$.
\endproof
Since strongly independent modulo $\mathfrak{a}$ depends on the leading terms of an ideal, we explore its connections with Gr\"obner basis. 
\begin{theorem}\label{grobnermaximal}
Let $\prec$ be a monomial ordering in $A[x_1, \ldots, x_n]$ and $S \subseteq X$ be a set of indeterminates.
Let $G$ be a Gr\"obner basis of an ideal, $\mathfrak{a} \subseteq A[x_1, \ldots, x_n]$ w.r.t. $\prec$. Then $S$ is strongly independent modulo $\mathfrak{a}$ w.r.t. $\prec$ 
if and only if $A[S] \cap \mathrm{lt}(G) = \emptyset$. 
\end{theorem}
\proof
If for some $g \in G$, $\mathrm{lt}(g) \in A[S]$, then $g \in A[S/(X\setminus S)] \cap \mathfrak{a}$ and therefore $S$ is not strongly independent modulo $\mathfrak{a}$. 
Conversely, assume there exists $f \in A[S/(X\setminus S)] \cap \mathfrak{a}$, then there exists at least one $g \in G$ such that $\mathrm{lm}(g) \mid \mathrm{lm}(f)$. 
Since $\mathrm{lt}(f) \in A[S]$, $\mathrm{lm}(g) \in A[S]$.  
\endproof
We can construct the LBS of $\mathfrak{a}$ w.r.t. $\prec$ from $G$ by the following algorithm which is analogous to \citep[Corollary 2.2]{Kredel:1988:maximalset}. 
\begin{corollary}
 Let $\prec$ be a monomial order in $A[x_1, \ldots,x_n]$ and $G$ be a Gr\"obner basis w.r.t. to $\prec$ for an ideal $\mathfrak{a}\subseteq A[x_1, \ldots, x_n]$. Algorithm~\ref{leftbasicsetcomputation} determines the left basic set 
 of $\mathfrak{a}$ w.r.t. $\prec$.
 \end{corollary}
 \begin{algorithm}
  \caption{Finding the Left Basic Set of an ideal $\mathfrak{a}$ in $A[x_1, \ldots, x_n]$} 
\begin{algorithmic}\label{leftbasicsetcomputation}
\STATE \textbf{Input} $G$, Gr\"obner basis of $\mathfrak{a} \subseteq A[x_1, \ldots, x_n]$ w.r.t. $\prec$ \\
\STATE \textbf{Output} $S$, Left Basic Set of $\mathfrak{a}$ w.r.t. $\prec$.
\STATE $S = \emptyset$, $U = \{x_1, \ldots, x_n\}$
\WHILE {$U \neq \emptyset$}
\STATE Select $x$ from $U$.
\STATE $U = U \setminus \{x\}$
\IF {$\mathrm{Mon}(A[S] \cup \{x\}) \cap \mathrm{lt}(G) = \emptyset$}
\STATE $S= S \cup \{x\}$
\ENDIF
\ENDWHILE
\end{algorithmic}
 \end{algorithm}

\begin{corollary}\label{dimensionlex}
Let $\mathfrak{a} \subseteq A[x_1, \ldots, x_n]$ be an ideal such that $A[x_1, \ldots, x_n]/\mathfrak{a}$ has a free $A$-module representation w.r.t. some lexicographic ordering, 
$\prec$ and $G$ be its monic short reduced Gr\"obner basis w.r.t.  $\prec$.  Let $S \subseteq X$ be a set of indeterminates such that
\begin{displaymath}
 \mathrm{Mon}(A[S]) \cap \mathrm{lt}(G) = \emptyset,
\end{displaymath}
and $S$ has the largest number of elements among all subsets of $X$ that satisfy the above equation. 
Then $S$ is maximal independent modulo $\mathfrak{a}$ and $|S| = \mathrm{cdim}(A[x_1, \ldots, x_n]/\mathfrak{a})$.
\end{corollary} 
\proof
This result is a direct consequence of Theorem~\ref{grobnermaximal} and Theorem~\ref{dimensionforideals}.
\endproof
The above result gives us an algorithmic technique to determine the combinatorial dimension of  $A[x_1, \ldots, x_n]/\mathfrak{a}$. 
It involves computing a Gr\"obner basis w.r.t. a lexicographic ordering.
Given a Noetherian integral domain $A$, we give below an explicit description of the algorithm to compute the combinatorial dimension of  $A$-algebras 
$A[x_1, \ldots, x_n]/\mathfrak{a}$, that have a free $A$-module representation w.r.t. a lexicographic ordering. 
The correctness of the algorithm follows from Corollary~\ref{dimensionlex}.
It consists of two routines, Algorithm~\ref{recursivelagorforcomb} and Algorithm~\ref{Dimensionrecursive}, the latter of which is recursive. Algorithm~\ref{Dimensionrecursive} also determines the maximal independent set of indeterminates modulo $\mathfrak{a}$.
This algorithm is  along the lines of the algorithm described in \citep[Section 3]{Kredel:1988:maximalset}. 
\begin{algorithm}[H]
  \caption{ Algorithm for finding the combinatorial dimension of $A[x_1, \ldots, x_n]/\mathfrak{a}$ for lexicographic orderings} 
\begin{algorithmic}\label{recursivelagorforcomb}
\STATE \textbf{Input} $G$, short reduced Gr\"obner basis of $\mathfrak{a} \subseteq A[x_1, \ldots, x_n]$ w.r.t. a lexicographic ordering $\prec$, \\
$X = \{x_1, \ldots, x_n\}$\\
\STATE \textbf{Output} $c$, combinatorial dimension of $A[x_1, \ldots, x_n]/\mathfrak{a}$\\
$\mathcal{S}$, the maximal set of indeterminates independent modulo $\mathfrak{a}$.
\IF {$G$ is not monic}
\STATE Exit
\ENDIF 
\STATE $c=0$, $S = \emptyset$, $U = X$, $\mathcal{M}=\emptyset$ \\
\COMMENT {Calls the recursive algorithm}
\STATE $\mathcal{M} = $ Algorithm~\ref{Dimensionrecursive}$(G, S,U,\mathcal{M})$
\STATE $\mathcal{S}=\mathcal{M}$
\WHILE {$\mathcal{M} \neq \emptyset$}
\STATE Select any $M$ from $\mathcal{M}$
\STATE $\mathcal{M} = \mathcal{M} \setminus\{ M \}$
\IF {$c\leq |M|$}
\STATE $c = |M| $
\ENDIF
\ENDWHILE
\end{algorithmic}
 \end{algorithm}

\begin{algorithm}[H]
  \caption{  Recursive algorithm for finding the maximal set of indeterminates independent modulo the ideal $\mathfrak{a} \subseteq A[x_1, \ldots, x_n]$ for lexicographic orderings} 
\begin{algorithmic}\label{Dimensionrecursive}
\STATE \textbf{Input} $G$, Gr\"obner basis of $\mathfrak{a} \subseteq A[x_1, \ldots, x_n]$ w.r.t. a lexicographic ordering $\prec$, \\
$S$, set of indeterminates such that $\mathrm{Mon}(S) \cap \mathrm{lt}(G) = \emptyset$, \\
$U$, a subset of the indeterminates set $X$,\\
$\mathcal{M}$, a set of already computed maximal sets $S^{'}$ with $\mathrm{Mon}(S^{'}) \cap \mathrm{lt}(G) = \emptyset$.  \\
\STATE \textbf{Output} $\mathcal{M}^{'}$, the updated set of maximal set of indeterminates $S^{'}$ with $\mathrm{Mon}(S^{'}) \cap \mathrm{lt}(G) = \emptyset$.  \\
\COMMENT{Finding the maximal independent sets of indeterminates}
\STATE $\mathcal{M}^{'} = \mathcal{M}$
\WHILE {$U\neq \emptyset$}
\STATE Select $u$ from $U$
\STATE $U = U\setminus \{u\}$
\IF {$\mathrm{Mon}(S \cup \{u\}) \cap \mathrm{lt} (G) = \emptyset$}
\STATE $\mathcal{M}^{'} = $ Algorithm~\ref{Dimensionrecursive}$(G, S \cup \{u\}, U, \mathcal{M}^{'})$
\ENDIF
\ENDWHILE \\
\COMMENT{ Testing if $S$ is already contained in some element of $\mathcal{M}^{'}$}
\STATE $\mathcal{M}^{''} = \mathcal{M}^{'}$, $ t = 1$
\WHILE {$\mathcal{M}^{''} \neq \emptyset$ and $t =1$}
\STATE Select $ M$ from $\mathcal{M}^{''}$, 
$\mathcal{M}^{''} = \mathcal{M}^{''} \setminus \{M \}$.
\IF {$S \subseteq  M$}
\STATE $t =0$
\ENDIF
\ENDWHILE
\IF {$t =1$}
\STATE $\mathcal{M}^{'}  = \mathcal{M}^{'} \cup \{S\}$
\ENDIF
\end{algorithmic}
 \end{algorithm}
The running time of the algorithm is exactly as that of computing the combinatorial dimension for fields except for the computation of short reduced Gr\"obner basis. 
The computation of short reduced Gr\"obner basis depends on the coefficient ring, $A$. 
When $A = \Bbbk$ or $ \mathbb{Z}$, the time complexity is doubly exponential (computation of a single Gr\"obner basis) and when $A = \Bbbk[y_1, \ldots, y_m]$, 
the complexity is still doubly exponential but involves two Gr\"obner basis computations, first in  $\Bbbk[y_1, \ldots, y_m]$ and then in $A[x_1, \ldots, x_n]$.
\section{Relation between Krull dimension  and  Combinatorial Dimension of $A[x_1, \ldots, x_n]/\mathfrak{a}$}
\label{dimension}
 The results we derive in this section will also help us derive a  relation between the degree of a Hilbert polynomial and Krull dimension (Section~\ref{krullwithhilbertsection}). 
\begin{lemma}\label{isomorphismtensor}
 Let $\mathfrak{a} \subseteq A[x_1, \ldots, x_n]$ be an ideal  and $M$ be an $A$-algebra.
 Then there exists a homomorphism from  $A[x_1, \ldots, x_n]$ to $M[x_1, \ldots, x_n]$. 
 Let $\mathfrak{a}^e$ represent the extension of the ideal $\mathfrak{a}$ in $M[x_1, \ldots, x_n]$ under the homomorphism.
We have, 
\begin{displaymath}
 M \otimes_{A} A[x_1, \ldots, x_n]/\mathfrak{a} \cong M[x_1, \ldots, x_n]/\mathfrak{a}^e.
\end{displaymath}
\end{lemma}
\proof
Clearly, $M[x_1, \ldots, x_n]/\mathfrak{a}^e$ is an $A$-module. Consider the following 
operation: for any $f + \mathfrak{a} \in A[x_1, \ldots, x_n]/\mathfrak{a}$ and $g + \mathfrak{a}^e \in M[x_1, \ldots, x_n]/\mathfrak{a}^e$, let  $(f+ \mathfrak{a})(g + \mathfrak{a}^e ) = fg + \mathfrak{a}^e$.
It is well defined because $\mathfrak{a} \subseteq \mathfrak{a}^e$. This implies $M[x_1, \ldots, x_n]/\mathfrak{a}^e$ is an $A[x_1, \ldots, x_n]/\mathfrak{a}$-module as well. 
We define the following homomorphism,
\begin{align*}
 \phi : A^{(A[x_1, \ldots, x_n]/\mathfrak{a} \times M)} &\rightarrow M[x_1, \ldots, x_n]/\mathfrak{a}^e \\
 \phi(\sum_{i \in \Lambda} (a_ix^{\alpha_i} + \mathfrak{a}, m_i)) & = \sum_{i \in \Lambda} (a_im_ix^{\alpha_i} + \mathfrak{a}^e).
\end{align*}
Note that $\phi$ is $A$-multilinear. Therefore, there exist the following homomorphisms,
\begin{displaymath}
 \psi: A[x_1, \ldots, x_n]/\mathfrak{a} \otimes_{A} M \rightarrow M[x_1, \ldots, x_n]/\mathfrak{a}^e 
\end{displaymath}
and 
\begin{displaymath}
 \pi:  A^{(A[x_1, \ldots, x_n]/\mathfrak{a} \times M)} \rightarrow A[x_1, \ldots, x_n]/\mathfrak{a} \otimes_{A} M
\end{displaymath}
such that $\phi = \psi \small\circ \pi$. 
Since $\phi$ is surjective, $\psi$ is surjective too. Consider, 
\begin{displaymath}
\psi ( \sum_{i \in \Lambda} (a_ix^{\alpha_i} + \mathfrak{a} \otimes_{A}   m_i)) = 0.
\end{displaymath}
We have,
\begin{displaymath}
 \sum_{i \in \Lambda} (a_ix^{\alpha_i} + \mathfrak{a} \otimes_{A}   m_i) = \sum_{i \in \Lambda} (x^{\alpha_i} + \mathfrak{a} \otimes_{A}   a_im_i).
 \end{displaymath}
Now, $\pi (\sum_{i \in \Lambda} (x^{\alpha_i} + \mathfrak{a},  a_im_i)) = \sum_{i \in \Lambda} (x^{\alpha_i} + \mathfrak{a} \otimes_{A}  a_im_i)$. This implies,
$\phi(\sum_{i \in \Lambda} (x^{\alpha_i} + \mathfrak{a},  a_im_i)) = 0$. 
Since $x^{\alpha_i}$s are standard monomials, if the sum is equal to zero then each $a_im_i = 0$.
Therefore, $ \sum_{i \in \Lambda} (a_ix^{\alpha_i} + \mathfrak{a} \otimes_{A}   m_i) = 0$ and $\psi$ is injective. We have the following isomorphism,
\begin{displaymath}
 M \otimes_{A} A[x_1, \ldots, x_n]/\mathfrak{a} \cong M[x_1, \ldots, x_n]/\mathfrak{a}^e. 
\end{displaymath}
\endproof
\begin{proposition}\label{gbidealandextensame}
Let $\mathfrak{a} \subseteq A[x_1, \ldots, x_n]$ be an ideal such that it has a monic short reduced Gr\"obner basis, $G = \{g_1, \ldots, g_t\}$ w.r.t. some monomial order $\prec$. 
Let $\mathfrak{p} \subsetneq A$ be a prime ideal and $k(\mathfrak{p})  (=A_\mathfrak{p}/\mathfrak{p}A_{\mathfrak{p}})$ be the residue  
field of $\mathfrak{p}$. Consider the ring homomorphism,
\begin{equation}\label{homophismresiduefield}
\nu : A[x_1, \ldots, x_n] \longrightarrow k(\mathfrak{p})[x_1, \ldots, x_n]
\end{equation}
such that $ \nu(x_i)  = x_i  \text{ for } x_i \in \{x_1, \ldots, x_n\}$ and for $a \in A$,
\[ \nu(a) =  \begin{cases} 
      0 &\text{ for } a \in \mathfrak{p} \\
      a & \text{ for } a \notin \mathfrak{p}.\\
     
   \end{cases}
\]
If $\mathfrak{a}^e$ is the extension of $\mathfrak{a}$ in $k(\mathfrak{p})[x_1, \ldots, x_n]$, then
$\nu(G) = \{\nu(g_1),\ldots, \nu(g_t)\}$ is a Gr\"obner basis for $\mathfrak{a}^e$. 
\end{proposition}
\proof
If
\begin{displaymath}
\langle\mathrm{lt}(\mathfrak{a})\rangle k(\mathfrak{p})[x_1, \ldots, x_n] = \langle\mathrm{lt}(\mathfrak{a}^e)\rangle,
\end{displaymath} 
 then 
$\nu(G) = \{\nu(g_1), \ldots, \nu(g_t)\}$ is a Gr\"obner basis of $\mathfrak{a}^e$ in $k(p)[x_1, \ldots, x_n]$. 
Since $G$ is a monic basis it also follows that $\nu(g), g \in G$ is monic  and therefore $\nu(G)$ is a monic Gr\"obner basis for $\mathfrak{a}^e$. 
We first show $\langle \mathrm{lt}(\mathfrak{a})\rangle k(\mathfrak{p})[x_1, \ldots, x_n] \subseteq \langle \mathrm{lt}(\mathfrak{a}^e)\rangle$. This is  true for any ring homomorphism 
\citep[Proposition 3.4]{Stillman:1991:Ringhomogrobner}. 
It is enough to show that each generator of $\langle\mathrm{lt}(\mathfrak{a})\rangle k(\mathfrak{p})[x_1, \ldots, x_n]$ belongs to $\langle \mathrm{lt}(\mathfrak{a}^e) \rangle$. 
The generators of  $\langle \mathrm{lt}(\mathfrak{a}) \rangle k(\mathfrak{p})[x_1, \ldots, x_n]$ are $\nu(\mathrm{lt}(f))$, $f \in \mathfrak{a}$. 
For each $f \in \mathfrak{a}$, either $\nu(\mathrm{lt}(f)) = 0$ if $\mathrm{lc}(f) \in \mathfrak{p}$ or $\nu(\mathrm{lt}(f)) = \mathrm{lt}(f) = \mathrm{lt}(\nu(f)) \in \langle \mathrm{lt}(\mathfrak{a}^e)\rangle$, 
if $\mathrm{lc}(f) \notin \mathfrak{p}$. 

Let $f \in \mathfrak{a}^e$ and $\mathrm{lt}(f) = cx^\alpha$. We have, 
\begin{displaymath}
f = \sum_{i=1}^{t} \nu(g_i)b_i, \hspace{5pt} b_i \in k(\mathfrak{p})[x_1, \ldots, x_n].
\end{displaymath}
We claim that $\mathrm{lt}(g_j) \mid x^\alpha$ for some $j \in \{1, \ldots, t\}$. If not,  for each $\mathrm{lt}(g_j)$,
$b_{j} =0$ since $G$ is a monic short reduced Gr\"obner basis and $\nu(\mathrm{lt}(g_i)) = \mathrm{lt}(g_i) = \mathrm{lm}(g_i)$.
Let $g_j \in G$ be such that $\mathrm{lm}(g_j) \mid x^\alpha$. Therefore, $x^\alpha \in \langle \mathrm{lt}(\mathfrak{a}) \rangle$ and  $cx^\alpha \in \langle \mathrm{lt}(\mathfrak{a}) \rangle k(\mathfrak{p})[x_1, \ldots, x_n]$. 
We have, $\nu(G)$ is a Gr\"obner basis for $\mathfrak{a}^e$. 
%
\endproof

\noindent Consider the ring homomorphism,
 \begin{equation}\label{simplering}
  f : A \longrightarrow A[x_1, \ldots, x_n]/\mathfrak{a}.
 \end{equation}
 We have the corresponding mapping associated with $f$, 
 \begin{equation} \label{specversion}
   f^{*}:\mathrm{Spec}(A[x_1, \ldots, x_n]/\mathfrak{a}) \longrightarrow  \mathrm{Spec}(A).
 \end{equation}
Consider a prime ideal $\mathfrak{p}$ in $A$. The subspace ${f^{*}}^{-1}(\mathfrak{p})$ of $\mathrm{Spec}(A[x_1, \ldots, x_n]/\mathfrak{a})$
is naturally homeomorphic to $\mathrm{Spec}(k(\mathfrak{p}) \otimes_A A[x_1, \ldots, x_n]/\mathfrak{a})$, where $k(\mathfrak{p})$ is the 
residue field of $\mathfrak{p}$, $A_\mathfrak{p}/\mathfrak{p}A_{\mathfrak{p}}$ \citep[Exercise 3.21]{Atiyah:1969:Commutativealgebra}. That is, we have a homeomorphism between the set of primes of 
$A[x_1, \ldots, x_n]/\mathfrak{a}$ lying over $\mathfrak{p}$ and  $\mathrm{Spec}(k(\mathfrak{p}) \otimes_A A[x_1, \ldots, x_n]/\mathfrak{a})$.
By Lemma~\ref{isomorphismtensor}, we have 
\begin{equation}\label{tensorpdtisomorphism}
k(\mathfrak{p})\otimes_{A} A[x_1, \ldots, x_n]/\mathfrak{a} \cong k(\mathfrak{p})[x_1, \ldots, x_n]/\mathfrak{a}^e.
\end{equation}
\begin{theorem}\label{matsumura}
Let $\mathfrak{a}$ be a proper ideal in $A[x_1, \ldots, x_n]$ such that it has a monic Gr\"obner basis w.r.t. some monomial ordering. Let $\mathfrak{p}$ be a prime ideal in $A$ and let $P$ be a prime ideal in $A[x_1, \ldots, x_n]/\mathfrak{a}$ such that $P$ is maximal among the prime ideals  lying over $\mathfrak{p}$. Then,
\begin{displaymath}
\mathrm{ht}(P) = \mathrm{ht}(\mathfrak{p}) + \mathrm{kdim}(k(\mathfrak{p})[x_1,\ldots, x_n]/\mathfrak{a}^e),
\end{displaymath}
where $k(\mathfrak{p})$ is the residue  
field of $\mathfrak{p}$  and $\mathfrak{a}^e$ is the extension of the ideal, $\mathfrak{a}$ under the ring homomorphism given by \eqref{homophismresiduefield}.
\end{theorem}
\proof
Consider the ring homomorphism given in \eqref{simplering},
\begin{displaymath}
f: A \longrightarrow A[x_1, \ldots, x_n]/\mathfrak{a}.
\end{displaymath}
Since $\mathfrak{a}$ has a monic Gr\"obner basis w.r.t. some monomial ordering, $A[x_1, \ldots, x_n]/\mathfrak{a}$ is a free $A$-module. This implies $f$ is a flat homomorphism of Noetherian rings and therefore we have from \citep[13.B Theorem 19]{Matsumura:1980:commutativealgebra}, 
\begin{displaymath}
\mathrm{ht}(P) = \mathrm{ht}(\mathfrak{p}) + \mathrm{kdim}((A[x_1, \ldots, x_n]/\mathfrak{a})_P \otimes k(\mathfrak{p})).
\end{displaymath}
To ease the notation, we denote 
$A[x_1, \ldots, x_n]/\mathfrak{a}$ as $\mathcal{A}$. 
The corresponding prime of $\mathcal{A} \otimes k(\mathfrak{p}) = \mathcal{A}_\mathfrak{p}/\mathfrak{p}\mathcal{A}_\mathfrak{p}$ is $P\mathcal{A}_\mathfrak{p}/\mathfrak{p}\mathcal{A}_\mathfrak{p}$. Let us denote this prime as $P^{*}$. Then by \citep[13.A]{Matsumura:1980:commutativealgebra} we have that the local ring, 
\begin{displaymath}
(\mathcal{A}\otimes k(\mathfrak{p}))_{P^{*}} = \mathcal{A}_P \otimes k(\mathfrak{p}). 
\end{displaymath}
Therefore,
\begin{displaymath}
\mathrm{kdim}(\mathcal{A}_P \otimes k(\mathfrak{p})) = \mathrm{kdim}((A[x_1, \ldots, x_n]/\mathfrak{a})_P \otimes k(\mathfrak{p}))=\mathrm{ht}(P^{*}).
\end{displaymath}
Consider $A[x_1, \ldots, x_n]/\mathfrak{a} \otimes k(\mathfrak{p}) $. By \eqref{tensorpdtisomorphism}, it is isomorphic to $k(\mathfrak{p})[x_1, \ldots, x_n]/\mathfrak{a}^e$. All maximal ideals in the affine algebra  $k(\mathfrak{p})[x_1, \ldots, x_n]/\mathfrak{a}^e$ are of the same height equal to $\mathrm{kdim}(k(\mathfrak{p})[x_1, \ldots, x_n]/\mathfrak{a}^e)$. Therefore, 
\begin{displaymath}
\mathrm{kdim}((A[x_1, \ldots, x_n]/\mathfrak{a})_P \otimes k(\mathfrak{p})) = \mathrm{kdim}(k(\mathfrak{p})[x_1, \ldots, x_n]/\mathfrak{a}^e),
\end{displaymath}
and we have,
\begin{displaymath}
\mathrm{ht}(P) = \mathrm{ht}(\mathfrak{p}) + \mathrm{kdim}(k(\mathfrak{p})[x_1,\ldots, x_n]/\mathfrak{a}^e).
\end{displaymath}
\endproof
\subsection{Krull dimension of $A$-algebras for lexicographic orderings}
\label{krulldimcombdimforlex}
 \begin{proposition}\label{samecdim}
Let $\mathfrak{a} \subseteq A[x_1, \ldots, x_n]$ be an ideal such that it has a monic short reduced Gr\"obner basis w.r.t. lexicographic ordering, $\prec$. 
Let $\mathfrak{p} \subsetneq A$ be a prime ideal and $k(\mathfrak{p})$ be the residue  
field of $\mathfrak{p}$ $ (=A_\mathfrak{p}/\mathfrak{p}A_{\mathfrak{p}})$. Let $\nu$ be the ring homomorphism as described in Proposition~\ref{gbidealandextensame} and 
 $\mathfrak{a}^e$ be the extension of $\mathfrak{a}$ in $k(\mathfrak{p})[x_1, \ldots, x_n]$. Then,
 \begin{displaymath}
  \mathrm{cdim}(k(\mathfrak{p})[x_1, \ldots, x_n]/\mathfrak{a}^e) = \mathrm{cdim}(A[x_1, \ldots, x_n]/\mathfrak{a}).
 \end{displaymath}
 \end{proposition}
 \proof
 Let $G$ be the monic short reduced Gr\"obner basis of $\mathfrak{a}$ w.r.t. a lexicographic ordering $\prec$.  From Proposition~\ref{gbidealandextensame}, we have that 
 $\nu(G)$ is a monic Gr\"obner basis for $\mathfrak{a}^e$ and $\mathrm{lt}(G) = \mathrm{lt}(\nu(G))$. Therefore, the set of indeterminates, $S \subseteq X$ 
 such that $\mathrm{Mon}(A[S]) \cap \mathrm{lt}(G) = \emptyset$ is the same as the set of indeterminates, $S^{'}\subseteq X$ that satisfy $\mathrm{Mon}(k(\mathfrak{p})[S^{'}]) \cap \mathrm{lt}(\nu(G)) = \emptyset$.
 Then by Corollary~\ref{dimensionlex}, 
 \begin{displaymath}
  \mathrm{cdim}(k(\mathfrak{p})[x_1, \ldots, x_n]/\mathfrak{a}^e) = \mathrm{cdim}(A[x_1, \ldots, x_n]/\mathfrak{a})
 \end{displaymath}
 and hence the proof. 
 \endproof
 \begin{corollary}\label{krullequalcombforlex}
   Let   $\mathfrak{a} \subseteq A[x_1, \ldots, x_n]$ be a proper ideal such that $A[x_1, \ldots, x_n]/\mathfrak{a}$ has a free $A$-module representation w.r.t.  
  a lexicographic order $\prec$.  
  Then,
  \begin{displaymath}
   \mathrm{kdim}(A[x_1, \ldots, x_n]/\mathfrak{a}) =  \mathrm{kdim}(A) +  \mathrm{cdim}(A[x_1, \ldots, x_n]/\mathfrak{a}).
  \end{displaymath}
 \end{corollary}
\proof
From Proposition~\ref{samecdim}, we have
\begin{displaymath}
  \mathrm{cdim}(k(\mathfrak{p})[x_1, \ldots, x_n]/\mathfrak{a}^e) = \mathrm{cdim}(A[x_1, \ldots, x_n]/\mathfrak{a}).
 \end{displaymath}
 When the coefficient ring is a field,
$ \mathrm{kdim}(k(\mathfrak{p})[x_1, \ldots, x_n]/\mathfrak{a}^e) =  \mathrm{cdim}(k(\mathfrak{p})[x_1, \ldots, x_n]/\mathfrak{a}^e)$. This implies that the equation in Proposition~\ref{matsumura} becomes,
\begin{displaymath}
\mathrm{ht}(P) = \mathrm{ht}(\mathfrak{p}) + \mathrm{cdim}(A[x_1,\ldots, x_n]/\mathfrak{a}).
\end{displaymath}
Since $\mathfrak{a}$ is a proper ideal with a monic Gr\"obner basis, the mapping in \eqref{specversion}, \\ $f^{*}:\mathrm{Spec}(A[x_1, \ldots, x_n]/\mathfrak{a}) \longrightarrow \mathrm{Spec}(A)$, is surjective and  we have,
\begin{displaymath}
\mathrm{kdim}(A[x_1, \ldots, x_n]/\mathfrak{a}) = \mathrm{kdim}(A) + \mathrm{cdim}(A[x_1, \ldots, x_n]/\mathfrak{a}).
\end{displaymath}
\endproof
Given a Noetherian integral domain,  using Corollary~\ref{krullequalcombforlex} we give a Gr\"obner basis algorithm to compute the Krull dimension of $A$-algebras, 
$A[x_1, \ldots, x_n]/\mathfrak{a}$ that have a free $A$-module representation w.r.t. a lexicographic ordering.  This is listed in Algorithm~\ref{recursivelagorforkrull}. 
This algorithm calls  Algorithm~\ref{recursivelagorforcomb},
which returns the maximal sets of indeterminates independent modulo $\mathfrak{a}$ and the combinatorial dimension of the corresponding $A$-algebra.
\begin{algorithm}[H]
  \caption{ Algorithm for finding the Krull dimension of $A[x_1, \ldots, x_n]/\mathfrak{a}$ for lexicographic orderings} 
\begin{algorithmic}\label{recursivelagorforkrull}
\STATE \textbf{Input} $G$, short reduced Gr\"obner basis of $\mathfrak{a} \subseteq A[x_1, \ldots, x_n]$ w.r.t. a lexicographic ordering, $\prec$, \\
$d_A$, Krull dimension of the ring, $A$, \\
$X = \{x_1, \ldots, x_n\}$\\
\STATE \textbf{Output} $d$, Krull dimension of $A[x_1, \ldots, x_n]/\mathfrak{a}$.
\IF {$G$ is not monic}
\STATE Exit
\ENDIF 
\STATE $c=0$, $S = \emptyset$, $t=0$, $\mathcal{S} = \emptyset$ \\
\COMMENT{Calls the combinatorial dimension algorithm}
\STATE $\mathcal{S},c = $ Algorithm~\ref{recursivelagorforcomb}$(G, X)$
\STATE $d = c+d_A$
\end{algorithmic}
 \end{algorithm}

\subsection{Examples}
\label{exampleslex}
We illustrate below examples that compute the Krull dimension of residue class rings of polynomial rings over a Noetherian integral domain, $A$ using combinatorial dimension.
\begin{example}
Consider the ideal $\mathfrak{a} = \langle xy, xz\rangle \subseteq A[x,y,z]$ and the lexicographic ordering $z\prec  y\prec x$.
Consider the $A$-algebra $\mathcal{A} = A[x,y,z]/\mathfrak{a}$. One way to determine the Krull dimension of $\mathcal{A}$ is given below.
We have, 
\begin{displaymath}
\mathrm{kdim} (\mathcal{A})=\mathrm{sup}\{\mathrm{kdim} (\mathcal{A}/\mathfrak{P}):\mathfrak{P} \text{ minimal prime}\}.
\end{displaymath} 
Let $\mathfrak{P}$ be a minimal prime of $\mathcal{A}$. Then $\mathfrak{P}=\mathfrak{p}/\langle xy,xz \rangle$ with $\mathfrak{p}$ prime in $A[x,y,z]$ and  minimal over $\langle xy,xz \rangle$. 
The associated isolated primes of  $\langle xy,xz \rangle$ are $\langle x \rangle$ and $\langle y,z \rangle$. 
Then,
\begin{align*}
\mathrm{kdim} (\mathcal{A}) & = \mathrm{sup} \{(\mathrm{kdim}(A[x,y,z]/\langle y,z \rangle), \mathrm{kdim}(A[x,y,z]/\langle x \rangle)\} \\ &= \mathrm{kdim}(A) + 2.
\end{align*}
We can also compute the Krull dimension using the relation we derived in the previous section. The short reduced  Gr\"obner basis of $\mathfrak{a}$
w.r.t. $\prec$ is $\{xy,xz\}$ and it is monic  and therefore $\mathcal{A}$ has a free $A$-module representation w.r.t. a lexicographic ordering. 
The $\mathrm{cdim}(\mathcal{A}) = 2$ since $S=\{y,z\}$ is a maximal independent set of indeterminates modulo $\mathfrak{a}$. 
Therefore we have,
\begin{displaymath}
\mathrm{kdim}(\mathcal{A}) = \mathrm{cdim}(\mathcal{A}) + \mathrm{kdim}(A) = \mathrm{kdim}(A) + 2.
\end{displaymath} 
\end{example}  
\begin{example}
Consider the ideal $\mathfrak{a} = \langle xy+1 \rangle \subseteq A[x,y]$ and the lexicographic ordering $ y\prec x$.
 One can see that the $A$-algebra $\mathcal{A} = A[x,y]/\mathfrak{a}$ is isomorphic to the ring of Laurent polynomials 
with coefficients in $A$, $A[x^{\pm 1}]$. Therefore, the Krull dimension of $\mathcal{A}$ is equal to $\mathrm{kdim}(A[x^{\pm 1}]) = \mathrm{kdim}(A) + 1$. 

We can use the relation we derived since $\mathcal{A}$ has a free $A$-module representation w.r.t. $\prec$. 
The $\mathrm{cdim}(\mathcal{A}) = 1$ with $S = \{x\}$ a maximal independent set modulo the ideal. 
Therefore $\mathrm{kdim}(\mathcal{A}) = \mathrm{kdim}(A) + 1$.
\end{example}

\begin{example}
 Let $\mathfrak{a} = \langle x^2y+ x + 1,  y^3+z+1 \rangle \subseteq A[x,y,z]$ be an ideal. 
 To determine the Krull dimension of the $A$-algebra $\mathcal{A} = A[x,y,z]/\mathfrak{a}$, 
we first compute the Gr\"obner basis of $\mathfrak{a}$ w.r.t. the lexicographic ordering $ z \prec y\prec x$. It is given by $\{y^3+z+1, x^2 z+x^2-xy^2-y^2, x^2y+x+1\}$.  
It is monic and therefore we can apply the relation we derived. We construct the Left Basic Set w.r.t. $\prec$, $S = \{z\}$. Therefore, 
$\mathrm{cdim}(\mathcal{A}) = |S| = 1$. Therefore, $\mathrm{kdim}(\mathcal{A}) = \mathrm{kdim}(A) + 1$. 
\end{example}

\begin{example}
 Let $\mathfrak{a} = \langle x^2+ 2x + 1,  y^3+2z+1 \rangle \subseteq A[x,y,z]$ be an ideal. 
The Gr\"obner basis of $\mathfrak{a}$ w.r.t. the lexicographic ordering $ z \prec y\prec x$ is $\{x^2+ 2x + 1,  y^3+2z+1 \}$.  
It is monic and therefore we can apply the relation we  derived to compute the Krull dimension of the $A$-algebra, $\mathcal{A} = A[x,y,z]/\mathfrak{a}$. The LBS w.r.t. $\prec$, $S = \{z\}$
and therefore, $\mathrm{cdim}(\mathcal{A}) = |S| = 1$ and $\mathrm{kdim}(\mathcal{A}) = \mathrm{kdim}(A) + 1$. 
\end{example}
\begin{example}
 Let $\mathfrak{a} = \langle x^2+ zx,  y+6z \rangle \subseteq \mathbb{Z}[x,y,z]$ be an ideal. 
 The Gr\"obner basis of $\mathfrak{a}$ w.r.t. the lexicographic ordering $ z \prec y\prec x$ is $\{x^2+ zx,  y+6z\}$.  
It is monic and therefore we can apply the relation we  derived to compute the Krull dimension of the $\mathbb{Z}$-algebra $ \mathbb{Z}[x,y,z]/\mathfrak{a}$. The LBS w.r.t. $\prec$, $S = \{z\}$ and 
therefore, $\mathrm{cdim}( \mathbb{Z}[x,y,z]/\mathfrak{a}) = |S| = 1$ and $\mathrm{kdim}( \mathbb{Z}[x,y,z]/\mathfrak{a}) = 2$. 
\end{example}
 \section{Hilbert Polynomials  in  $A[x_1, \ldots, x_n]$}
  \label{Hilbert}
 \subsection{Hilbert function and Hilbert series}

\begin{proposition}\label{basis}
Let $\mathfrak{a}\subseteq A[x_1,\ldots,x_n]$ be an ideal such that it has  a monic short reduced Gr\"obner basis, $G = \{g_1, \ldots, g_t\}$  
w.r.t. a degree compatible monomial ordering. We denote $\mathcal{A} = A[x_1,\ldots,x_n]/\mathfrak{a}$ . 
For $d$ a nonnegative
integer, we define 
\begin{displaymath}
\mathcal{A}_{\leq d} = \{f+\mathfrak{a} : f \in A[x_1,\ldots, x_n], \mathrm{deg}(f) \leq d\}.
\end{displaymath} 
Then, $\mathcal{A}_{\leq d} $ is a finitely generated, free $A$-module.
\end{proposition}
\proof
Let a basis for $\mathcal{A}$ be given by the set, $\mathcal{B} = \{ x^\alpha + \mathfrak{a} : \mathrm{lm}(g_i) \nmid x^\alpha \}$. 
Consider the following set, $\mathcal{B}^{(d)} = \{x^\alpha + \mathfrak{a} : x^\alpha + \mathfrak{a}  \in \mathcal{B}, \mathrm{deg}(x^\alpha) \leq d\}$.
\begin{claim*}
$\mathcal{B}^{(d)} $ is an $A$-module basis for $\mathcal{A}_{\leq d} $.
\end{claim*}
Clearly, $\mathcal{B}^{(d)} $ is a subset of $\mathcal{A}_{\leq d} $.
Consider $f + \mathfrak{a} \in \mathcal{A}_{\leq d} $.  Since  $\mathrm{deg}(f) \leq d$ and we have a degree compatible ordering, $\mathrm{lt}(f) \leq d$. This implies that $f + \mathfrak{a}$ 
can be written as $\sum_{x^{\alpha}+\mathfrak{a} \in \mathcal{B}^{(d)}} a_i(x^\alpha+\mathfrak{a})$, $a_i \in A$. Thus, $\mathcal{B}^{(d)} $ generates $\mathcal{A}_{\leq d} $. $\mathcal{B}^{(d)} $ 
is linearly independent since it is a subset of the basis, $\mathcal{B}$. We have, therefore, that $\mathcal{A}_{\leq d} $ is free and finitely generated.
\endproof
We refer to the size of $\mathcal{B}^{(d)}$ as the free rank of  $\mathcal{A}_{\leq d} $ and it is denoted as $\mathrm{FreeRank}_{A}(\mathcal{A}_{\leq d})$. 
Note that any two bases for a free module over a commutative ring  have the same cardinality.

Consider $\mathcal{A} =  A[x_1,\ldots,x_n]/\mathfrak{a}$ such that it has a free $A$-module 
representation w.r.t. a degree compatible monomial ordering.  We define the Hilbert function, 
$h_\mathfrak{a} : \mathbb{Z}_{\geq 0} \rightarrow \mathbb{Z}_{\geq 0}$ as 
\begin{displaymath}
 h_\mathfrak{a}(d) = \mathrm{FreeRank}_{A}(\mathcal{A}_{\leq d}).
\end{displaymath}
The formal power series 
\begin{displaymath}
 H_\mathfrak{a}(t) = \sum_{d=0}^\infty h_\mathfrak{a}(d)t^d \in \mathbb{Z}[[t]]
\end{displaymath}
is called the Hilbert series of $\mathfrak{a}$.

\begin{theorem}\label{leadingtermhilbert}
 Let $\mathfrak{a} \subseteq A[x_1,\ldots,x_n]$ be an ideal such that $ A[x_1,\ldots,x_n]/\mathfrak{a}$ has a free $A$-module representation w.r.t. a degree compatible ordering. Then,
 \begin{displaymath}
  H_\mathfrak{a}(t) = H_{\langle \mathrm{lt}(\mathfrak{a})\rangle}(t).
 \end{displaymath}
\end{theorem}
\proof
 Let $\mathcal{A} = A[x_1,\ldots,x_n]/\mathfrak{a}$ and $G= \{g_1,\ldots,g_t\}$ be a short reduced Gr\"obner basis of $\mathfrak{a}$ w.r.t a degree compatible ordering.
 Consider the following map for a specific set of coset representatives $C_{J_{x^\alpha}}$, $x^\alpha \in \mathrm{Mon}(A[x_1, \ldots, x_n])$, in $A$. 
 \begin{align*}
  \phi:\mathcal{A} &\longrightarrow A[x_1,\ldots,x_n]\\
  g+\mathfrak{a} &\longmapsto\eta_G(g).
 \end{align*}
The map is well defined \citep[Lemma 4.3.3.]{Adams:1994:introtogrobnerbasis}. 
For every $d \in \mathbb{Z}_{\geq 0}$, we have the restriction map,
\begin{displaymath}
 \phi_{d}:\mathcal{A}_{\leq d} \rightarrow A[x_1,\ldots,x_n].
\end{displaymath}
Let $V_{d}\subseteq A[x_1,\ldots,x_n]$ be the submodule spanned by all the monomials $t$ with degree $\leq d$ and $t \notin \langle \mathrm{lt}(\mathfrak{a}) \rangle$.
Since all $f \in V_d$ are in the normal form w.r.t. $G$, we get $f = \eta_G(f) = \phi_d(f+\mathfrak{a})$. 
Therefore, $V_d \subseteq im(\phi_d)$, the image of $\phi_d$. Let $f \in im(\phi_d)$. This implies $f = \eta_G(g)$ for some polynomial $g \in A[x_1,\ldots,x_n]$ and 
$\mathrm{Mon}(f)\notin \langle \mathrm{lt}(\mathfrak{a}) \rangle$. 
We have that the degree of each monomial in $f$ is less than $d$ since the ordering is degree compatible. Therefore, $f \in V_d$ and $h_\mathfrak{a}(d) = \mathrm{FreeRank}(V_d)$. 
Note that the definition of $V_d$ depends only on the leading term ideal and therefore two ideals with the same leading term ideal will have the same Hilbert series.
\endproof
Algorithm~\ref{hilbertserre} gives a Gr\"obner basis method to calculate the Hilbert series of an ideal in $A[x_1, \ldots, x_n]$.  
\begin{algorithm}
\caption{Computing the Hilbert series of an ideal $\mathfrak{a}$ in  $A[x_1, \ldots, x_n]$ when $A[x_1,\ldots,x_n]/\mathfrak{a}$  has a  free $A$-module representation w.r.t. a degree 
compatible monomial ordering.} 
\begin{algorithmic}
\label{hilbertserre}
\STATE \textbf{Input} 
A degree compatible monomial ordering $\prec$,\\
$G=\{g_1,\ldots,g_s\}$, a monic short reduced Gr\"obner basis of $\mathfrak{a}$ based on the ordering, $\prec$.
\STATE \textbf{Output} Hilbert series $H_{\mathfrak{a}}(t)$.
\STATE Let $m_1,\ldots,m_s$ be the leading monomials of $G$.
\IF {$s=0$}
\STATE  Return $H_\mathfrak{a}(t) = \frac{1}{{(1-t)}^{n+1}}$.
\ELSE
\STATE $J=\langle m_2, \cdots, m_s\rangle$ and \\ $J^{'}=\langle\mathrm{lcm}(m_1,m_2),\cdots,\mathrm{lcm}(m_1,m_s)\rangle$.
\STATE Compute $H_J(t)$ and $H_{J^{'}}(t)$ by a recursive call of the algorithm.
\STATE Return 
\begin{displaymath}
 H_\mathfrak{a}(t) = \frac{1-t^{\mathrm{deg}(m_1)}}{{(1-t)}^{n+1}} + H_J(t) - H_{J^{'}}(t). 
\end{displaymath}
\ENDIF
\end{algorithmic}
\end{algorithm}
\begin{proposition}
Algorithm~\ref{hilbertserre} terminates after finitely many steps and calculates $H_{\mathfrak{a}}(t)$ correctly. 
\end{proposition}
\proof
With the assumption that $A[x_1, \ldots, x_n]/\mathfrak{a}$ has a free $A$-module representation w.r.t. a degree compatible monomial ordering, the proof is identical to that of fields \citep[Theorem 11.9]{Kemper:2011:hilbertpolynomialGB}.
\endproof
The Hilbert-Serre theorem follows as a natural consequence of the above algorithm.
\begin{theorem}[Hilbert-Serre theorem]
\label{hilbert-serre}
Let  $\mathfrak{a} \subseteq A[x_1, \ldots, x_n]$ be an ideal such that 
$A[x_1,\ldots,x_n]/\mathfrak{a}$ has a free $A$-module representation w.r.t. a degree compatible ordering. Then the Hilbert series of the ideal has the form,
\begin{displaymath}
 H_\mathfrak{a}(t) = \frac{a_0 + a_1t + \cdots + a_k t^k}{(1-t)^{n+1}},
\end{displaymath}
with $k\in \mathbb{Z}_{\geq 0}$ and $a_i \in \mathbb{Z}$. 
Moreover, the Hilbert function $h_\mathfrak{a}(d)$ is a polynomial 
for large $d$. The polynomial,
\begin{displaymath}
 p_\mathfrak{a} = \sum_{i=0}^k a_i C(x+n-i,n) \in \mathbb{Q}[x]
\end{displaymath}
called the Hilbert polynomial satisfies $h_\mathfrak{a}(d) = p_\mathfrak{a}(d)$ for sufficiently large integer $d$. 
 \end{theorem}
Whenever the $A$-module $A[x_1, \ldots, x_n]/\mathfrak{a}$ has a free $A$-module representation w.r.t. a degree compatible monomial ordering
all the properties of Hilbert functions for affine $\Bbbk$-algebras hold here as well. 
\subsection{Relation between Hilbert polynomials and combinatorial dimension}
\label{hilbertequalscomb}
  Let $\mathfrak{a}\subseteq A[x_1, \ldots, x_n]$ be an ideal such that $A[x_1, \ldots, x_n]/\mathfrak{a}$ has a free $A$-module representation w.r.t. 
a degree compatible monomial ordering. We first show that the degree of the Hilbert polynomial  is equal to its combinatorial dimension. 
A free $A$-module representation implies monic leading terms and  
this implies all the properties of Hilbert functions for leading term ideals follow exactly as that of fields. 
%
One such property is the equivalence of the combinatorial dimension of $A[x_1, \ldots, x_n]/\langle \mathrm{lt}(\mathfrak{a})\rangle$ and the degree of Hilbert polynomial of $\langle \mathrm{lt}(\mathfrak{a})\rangle $. 
\begin{theorem}\label{hilbertequaldimmonomial}
If $\mathfrak{a} \subseteq A[x_1, \ldots, x_n]$ is an ideal such that $A[x_1, \ldots, x_n]/\mathfrak{a}$ has a free $A$-module representation w.r.t. a degree compatible monomial order, then the degree of the Hilbert polynomial of $\langle\mathrm{lt}(\mathfrak{a})\rangle$ is equal to the combinatorial dimension of $A[x_1, \ldots, x_n]/\langle\mathrm{lt}(\mathfrak{a})\rangle$.                                                                                                            
\end{theorem}
We will now show that for any arbitrary ideal $\mathfrak{a}$, $\mathrm{cdim}(A[x_1, \ldots, x_n]/\mathfrak{a})$ is equal to the degree of the Hilbert polynomial of $\mathfrak{a}$.
\begin{theorem}\label{cdimequalsdeghilbert}
Let $\mathfrak{a} \subseteq A[x_1, \ldots, x_n]$ be a proper ideal 
such that $A[x_1, \ldots, x_n]/\mathfrak{a}$ has a free $A$-module representation w.r.t. a degree compatible ordering $\prec$. Then, $\mathrm{cdim}(A[x_1, \ldots, x_n]/\mathfrak{a})$ 
equals  the degree of the Hilbert polynomial of $\mathfrak{a}$.
\end{theorem}
\proof
Let $d$ denote the combinatorial dimension of $\mathfrak{a}$.  Let the set $\{x_{i_1}, \ldots, x_{i_d}\}$ be a set of independent indeterminates modulo
$\mathfrak{a}$ of maximal cardinality.  Let $s$ be a non-negative integer. From Theorem~\ref{basis}, we have that $\mathrm{Mon}( A[x_{i_1}, \ldots, x_{i_d}])_{\leq s}$ is a linearly independent set of 
$\mathcal{A}_{\leq s} $. 
Therefore, $C(d+s, s) \leq h_\mathfrak{a}(s)$. Since the binomial coefficient is a polynomial function in $s$ of degree $d$, 
the $\mathrm{cdim}(A[x_1, \ldots, x_n]/\mathfrak{a})$ is at most the degree of the Hilbert polynomial. 

Let $\langle \mathrm{lt}(\mathfrak{a}) \rangle$ be the leading term ideal of $\mathfrak{a}$ w.r.t. $\prec$. If $S = \{x_{i_1}, \ldots, x_{i_k}\} \subseteq \{x_1, \ldots, x_n\}$ is not
independent modulo $\mathfrak{a}$, then there exists a non-zero polynomial $f \in \mathfrak{a} \cap A[x_{i_1}, \ldots, x_{i_k}]$. 
We have, $\mathrm{lm}(f) \in \langle \mathrm{lt}(\mathfrak{a}) \rangle \cap A[x_{i_1}, \ldots, x_{i_k}]$. This implies $S$ is not independent modulo $\langle \mathrm{lt}(\mathfrak{a}) \rangle$.
Therefore, the set of independent indeterminates modulo $\langle \mathrm{lt}(\mathfrak{a})\rangle$ is a subset of the set of independent indeterminates modulo $\mathfrak{a}$. Therefore,
$\mathrm{cdim}(A[x_1, \ldots, x_n]/\langle \mathrm{lt}(\mathfrak{a}) \rangle) \leq \mathrm{cdim}(A[x_1, \ldots, x_n]/\mathfrak{a})$. By Theorem~\ref{leadingtermhilbert} and Theorem~\ref{hilbertequaldimmonomial}, 
we have that the degree of the Hilbert polynomial is at most $\mathrm{cdim}(A[x_1, \ldots, x_n]/\mathfrak{a})$.
\endproof
 This corollary directly follows.
 \begin{corollary} \label{leadingtermcorollary}
 Given a proper ideal $\mathfrak{a} \subseteq A[x_1, \ldots, x_n]$ such that $A[x_1, \ldots, x_n]/\mathfrak{a}$ has a free $A$-module representation w.r.t. a degree compatible ordering.
 If 
 $S$ is a set of maximal cardinality of indeterminates that are independent modulo $\langle\mathrm{lt}(\mathfrak{a})\rangle$, 
 then $S$ is a set of maximal cardinality of indeterminates that are independent modulo $\mathfrak{a}$. 
 Also,
 \begin{displaymath}
  \mathrm{cdim}(A[x_1, \ldots, x_n]/\mathfrak{a}) = \mathrm{cdim}(A[x_1, \ldots, x_n]/\langle\mathrm{lt}(\mathfrak{a})\rangle). 
 \end{displaymath}
 \end{corollary}
 \subsection{Krull dimension of $A$-algebras for degree compatible orderings}
 \label{krullwithhilbertsection}
  \begin{proposition}\label{samehilbertdegreeextension}
Let $\mathfrak{a} \subseteq A[x_1, \ldots, x_n]$ be an ideal such that it has a monic short reduced Gr\"obner basis w.r.t. a degree compatible ordering $\prec$. 
Let $\mathfrak{p} \subsetneq A$ be a prime ideal and $k(\mathfrak{p})$ be the residue  
field of $\mathfrak{p}$ $ (=A_\mathfrak{p}/\mathfrak{p}A_{\mathfrak{p}})$. Let $\nu$ be the ring homomorphism as described in Proposition~\ref{gbidealandextensame} and 
 $\mathfrak{a}^e$ be the extension of $\mathfrak{a}$ in $k(\mathfrak{p})[x_1, \ldots, x_n]$. Then,
 \begin{displaymath}
 H_{\mathfrak{a}^e} (t) = H_{\mathfrak{a}}(t).
 \end{displaymath}
 \end{proposition}
 \proof
 Let $G$ be the monic short reduced Gr\"obner basis of $\mathfrak{a}$ w.r.t. a lexicographic ordering $\prec$.  From Proposition~\ref{gbidealandextensame}, we have that 
 $\nu(G)$ is a monic Gr\"obner basis for $\mathfrak{a}^e$ and $\mathrm{lt}(G) = \mathrm{lt}(\nu(G))$. Therefore, 
 we have  $H_{\mathfrak{a}^e} (t) = H_{\langle \mathrm{lt}(\mathfrak{a})\rangle} (t)$. From Theorem~\ref{leadingtermhilbert} we have, $H_{\mathfrak{a}^e} (t) = H_{\langle \mathrm{lt}(\mathfrak{a})\rangle} (t) =H_{\mathfrak{a}}(t)$. 
 \endproof
In the case of $A$-algebras with a free $A$-module representation w.r.t. a lexicographic ordering, we have seen that 
$\mathrm{cdim}(k(\mathfrak{p})[x_1, \ldots, x_n]/\mathfrak{a}^e) = \mathrm{cdim}(A[x_1, \ldots, x_n]/\mathfrak{a})$ (Proposition~\ref{samecdim}). 
This is true in  the case of $A$-algebras with a free $A$-module representation w.r.t. a degree compatible monomial ordering as well.  
 \begin{proposition}\label{samecdimdegcomp}
Let $\mathfrak{a} \subseteq A[x_1, \ldots, x_n]$ be an ideal such that it has a monic short reduced Gr\"obner basis w.r.t. a degree compatible ordering, $\prec$. 
Let $\mathfrak{p} \subsetneq A$ be a prime ideal and $k(\mathfrak{p})$ be the residue  
field of $\mathfrak{p}$ $ (=A_\mathfrak{p}/\mathfrak{p}A_{\mathfrak{p}})$. Let $\nu$ be the ring homomorphism as described in Proposition~\ref{gbidealandextensame} and 
 $\mathfrak{a}^e$ be the extension of $\mathfrak{a}$ in $k(\mathfrak{p})[x_1, \ldots, x_n]$. Then,
 \begin{displaymath}
  \mathrm{cdim}(k(\mathfrak{p})[x_1, \ldots, x_n]/\mathfrak{a}^e) = \mathrm{cdim}(A[x_1, \ldots, x_n]/\mathfrak{a}).
 \end{displaymath}
 \end{proposition}
 \proof
 Let $G$ be the monic short reduced Gr\"obner basis of $\mathfrak{a}$ w.r.t. $\prec$.  As shown previously,  
 $\nu(G)$ is a monic Gr\"obner basis for $\mathfrak{a}^e$ and $\mathrm{lt}(G) = \mathrm{lt}(\nu(G))$. From Proposition~\ref{samehilbertdegreeextension} we have $\mathrm{deg}(p_{\mathfrak{a}^e}) = \mathrm{deg}(p_\mathfrak{a})$. From Theorem~\ref{cdimequalsdeghilbert}, $\mathrm{cdim}(A[x_1, \ldots, x_n]/\mathfrak{a}) = \mathrm{deg}(p_\mathfrak{a})$. 
  Since over fields, $\mathrm{cdim}(k(\mathfrak{p})[x_1, \ldots, x_n]/\mathfrak{a}^e) = \mathrm{deg}(p_{\mathfrak{a}^e})$, we have 
 \begin{displaymath}
  \mathrm{cdim}(k(\mathfrak{p})[x_1, \ldots, x_n]/\mathfrak{a}^e) = \mathrm{cdim}(A[x_1, \ldots, x_n]/\mathfrak{a}).
 \end{displaymath}
 \endproof
%
\begin{corollary}\label{krullequalcombfordeg}
Let $A[x_1, \ldots, x_n]/\mathfrak{a}$ be a finitely generated $A$-algebra such that it has a free $A$-module representation w.r.t.  
  a degree compatible ordering $\prec$.  
  Then,
 \begin{align*}
  \mathrm{kdim}(A[x_1, \ldots, x_n]/\mathfrak{a}) &=  \mathrm{kdim}(A) +  \mathrm{cdim}(A[x_1, \ldots, x_n]/\mathfrak{a}) \\
 & = \mathrm{kdim}(A) + \mathrm{deg}(p_\mathfrak{a}).
 \end{align*}
 \end{corollary}
\proof
The proof goes along the same lines as Proposition~\ref{samecdim}. 
From Proposition~\ref{samecdimdegcomp}, we have
\begin{displaymath}
  \mathrm{cdim}(k(\mathfrak{p})[x_1, \ldots, x_n]/\mathfrak{a}^e) = \mathrm{cdim}(A[x_1, \ldots, x_n]/\mathfrak{a}).
 \end{displaymath}
 When the coefficient ring is a field,
$ \mathrm{kdim}(k(\mathfrak{p})[x_1, \ldots, x_n]/\mathfrak{a}^e) =  \mathrm{cdim}(k(\mathfrak{p})[x_1, \ldots, x_n]/\mathfrak{a}^e)$. This implies that the equation in Proposition~\ref{matsumura} becomes,
\begin{displaymath}
\mathrm{ht}(P) = \mathrm{ht}(\mathfrak{p}) + \mathrm{cdim}(A[x_1,\ldots, x_n]/\mathfrak{a}).
\end{displaymath}
Since $\mathfrak{a}$ is a proper ideal with a monic Gr\"obner basis, the mapping in \eqref{specversion},\\  $f^{*}:\mathrm{Spec}(A[x_1, \ldots, x_n]/\mathfrak{a}) \longrightarrow \mathrm{Spec}(A)$, is surjective and  we have,
\begin{displaymath}
\mathrm{kdim}(A[x_1, \ldots, x_n]/\mathfrak{a}) = \mathrm{kdim}(A) + \mathrm{cdim}(A[x_1, \ldots, x_n]/\mathfrak{a}).
\end{displaymath}
Since by Theorem~\ref{cdimequalsdeghilbert}, $\mathrm{deg}(p_\mathfrak{a}) = \mathrm{cdim}(A[x_1, \ldots, x_n]/\mathfrak{a})$, we have the result.
\endproof
We give below an algorithm (Algorithm~\ref{recursivelagorforkrulldeg}) to compute the Krull dimension  of  certain $A$-algebras, 
$A[x_1, \ldots, x_n]/\mathfrak{a}$, that have a free $A$-module representation w.r.t. a degree compatible ordering. 
The correctness of the algorithm follows from Corollary~\ref{krullequalcombfordeg}.
\begin{algorithm}[H]
  \caption{ Algorithm for finding the Krull dimension of $A[x_1, \ldots, x_n]/\mathfrak{a}$ for degree compatible orderings} 
\begin{algorithmic}\label{recursivelagorforkrulldeg}
\STATE \textbf{Input} $G$, short reduced Gr\"obner basis of $\mathfrak{a} \subseteq A[x_1, \ldots, x_n]$ w.r.t. a degree compatible monomial ordering, $\prec$, \\
$d_A$, Krull dimension of  $A$, \\
\STATE \textbf{Output} $d$, Krull dimension of $A[x_1, \ldots, x_n]/\mathfrak{a}$.
\IF {$G$ is not monic}
\STATE Exit
\ENDIF \\
\COMMENT{Calls the Hilbert Serre algorithm}
\STATE $H_\mathfrak{a} (t)= $ Algorithm~\ref{hilbertserre}$(G, \prec)$\\
 \COMMENT{ $H_\mathfrak{a}(t) $ is of the form $\frac{a_0 + a_1t + \cdots + a_k t^k}{(1-t)^{n+1}}$}
\STATE $ p_\mathfrak{a} (x) = \sum_{i=0}^k a_i C(x+n-i,n)$
\STATE $k = \mathrm{deg}(p_\mathfrak{a})$
\STATE $d = k+d_A$
\end{algorithmic}
 \end{algorithm}

 \subsection{Examples}
 We give below examples that compute the Krull dimension of residue class rings of polynomial rings over a Noetherian integral domain $A$ using Hilbert polynomials.
\begin{example}
Consider the ideal $\mathfrak{a} = \langle xy, xz\rangle \subseteq A[x,y,z]$ and the deglex ordering with $z\prec  y\prec x$.
Consider the $A$-algebra $\mathcal{A} = A[x,y,z]/\mathfrak{a}$. 
 The short reduced  Gr\"obner basis of $\mathfrak{a}$
w.r.t. $\prec$  is $\{xy,xz\}$ and it is monic.  Therefore $\mathcal{A}$ has a free $A$-module representation w.r.t. a degree compatible monomial ordering. By using the recursive algorithm Algorithm~\ref{hilbertserre}, we have 
\begin{align*}
H_\mathfrak{a}(t) &= \frac{-t^2+t+1}{(1-t)^3}\\
&= 1+ 4t+ 8t^2 + 13 t^3 + \cdots \\
p_\mathfrak{a}(x) &=   x^2 + 5x + 2.\\
\mathrm{deg}(p_\mathfrak{a}) &= 2.
\end{align*}
Using Corollary~\ref{krullequalcombfordeg},  we have,
\begin{displaymath}
\mathrm{kdim}(\mathcal{A}) = \mathrm{deg}(p_\mathfrak{a})+ \mathrm{kdim}(A) = \mathrm{kdim}(A) + 2.
\end{displaymath} 
\end{example}

\begin{example}
Consider the ideal $\mathfrak{a} = \langle xy+1 \rangle \subseteq A[x,y]$ and deglex ordering with $ y\prec x$.
We determine below the Krull dimension of the $A$-algebra $\mathcal{A} = A[x,y]/\mathfrak{a}$.  
We have 
\begin{align*}
H_\mathfrak{a}(t) &= \frac{1-t^2}{(1-t)^3}\\
&= 1+ 3t+ 5t^2 + 7 t^3 +9t^4+ \cdots \\
p_\mathfrak{a}(x) &=   2x + 1.\\
\mathrm{deg}(p_\mathfrak{a}) &= 1.
\end{align*}
Therefore, $\mathrm{kdim}(\mathcal{A}) = \mathrm{kdim}(A) + 1$.
\end{example}
\begin{example}
Let $\mathfrak{a} = \langle x^2+ zx,  y+6z \rangle \subseteq \mathbb{Z}[x,y,z]$ be an ideal. 
 The Gr\"obner basis of $\mathfrak{a}$ w.r.t. the deglex ordering $ z \prec y\prec x$ is $\{x^2+ zx,  y+6z\}$. 
 We have 
\begin{align*}
H_\mathfrak{a}(t) &= \frac{t^3-t^2-t+1}{(1-t)^4}\\
&= 1+ 3t+ 5t^2 + 7 t^3 + \cdots \\
p_\mathfrak{a}(x) &=   2x + 1.\\
\mathrm{deg}(p_\mathfrak{a}) &= 1.
\end{align*} 
Therefore, $\mathrm{kdim}( \mathbb{Z}[x,y,z]/\mathfrak{a}) = \mathrm{kdim}(\mathbb{Z}) + 1 = 2$. 
\end{example}
 \section*{Concluding Remarks}
 As we can see from the examples given in this paper, to determine the Krull dimension of $A[x_1, \ldots, x_n]/\mathfrak{a}$,
previously, one had to exploit the individual properties of each
ideal. In this paper, we derived a relation
between combinatorial dimension and Krull dimension that gives us
an algorithmic method to compute the Krull dimension of the $A$-algebra provided 
 it has a free $A$-module representation w.r.t. either a lexicographic or degree compatible monomial order. 
 A natural question to ask is can we have the similar relation for other monomial orders. For polynomial rings over fields, 
 the relation for all monomial orders is proved using \citep[Theorem 3.1]{Carra:1987:allorderings}.  
 An affirmative answer seems likely for $A[x_1, \ldots, x_n]$ as well but we have yet to have a formal proof. 
 
 In \citep{Kredel:1988:maximalset}, the authors conjecture that for a prime ideal $\mathfrak{a}\subseteq \Bbbk[x_1, \ldots,x_n]$ any maximal set of indeterminates strongly independent mod $\mathfrak{a}$ 
 is also maximal independent mod
 $\mathfrak{a}$ and hence determines the  dimension of $\Bbbk[x_1, \ldots,x_n]/\mathfrak{a}$. 
 The conjecture was shown to be true in \citep{Strumfels:1995:provingconjecture}. We conjecture the same for prime ideals in $A[x_1, \ldots, x_n]$.
 In this paper, for a Noetherian integral domain $A$ we have shown that the maximal strongly independent set of indeterminates constructed from the left basic set w.r.t. a lexicographic ordering is also maximal 
 independent mod $\mathfrak{a}$ and
 equal to the combinatorial dimension of $A[x_1, \ldots,x_n]/\mathfrak{a}$, when $A[x_1, \ldots,x_n]/\mathfrak{a}$ has a free $A$-module representation w.r.t. the ordering. 
 We conjecture that for a prime ideal  $\mathfrak{a}\subseteq A[x_1, \ldots,x_n]$, any maximal set of 
 indeterminates strongly independent mod $\mathfrak{a}$ is also maximal independent mod
 $\mathfrak{a}$. If this is true, then the cardinality of such a set determines the combinatorial and 
 Krull dimension of $A[x_1, \ldots,x_n]/\mathfrak{a}$.

 

{\footnotesize

\section*{Acknowledgments}
The authors thank the anonymous reviewers for suggestions and comments
that have let to a significant improvement of the manuscript.
  

}


\end{document}